\documentclass[a4paper,12pt]{article}
\usepackage{jheppub} % for details on the use of the package, please see the JINST-author-manual
\usepackage{subfigure}
\usepackage{graphicx}
\usepackage{color}
 
\usepackage{soul}

\makeatletter
\gdef\@fpheader{Accepted in JHEP\\
\footnotesize{DOI: 10.1007/JHEP10(2023)189}}
\makeatother

\title{\boldmath Circular string in a Black p-brane leading to chaos}

\author{Pinaki Dutta,}
\author{Kamal L. Panigrahi}
\author{and Balbeer Singh}
\affiliation{Department of Physics, Indian Institute of Technology Kharagpur,\\ Kharagpur 721 302, India}
\emailAdd{coolguddu0815@kgpian.iitkgp.ac.in}
\emailAdd{panigrahi@phy.iitkgp.ac.in}
\emailAdd{curiosity1729@kgpian.iitkgp.ac.in}
\abstract{We consider a pulsating string near a  non-extremal black p-brane (p=5 and p=6) and investigate the chaos in the corresponding string dynamics by examining the Fast Lyapunov indicator(FLI) and Poincare section. In our system, the energy and the charge play the role of control parameters. For generic values of these parameters, the numerical results show that the dynamics primarily fall into three modes: capture, escape to infinity, and quasiperiodic depending on the initial location (near to or far away from the black brane horizon) of the string.
Finally, probing  for different values of the winding number $(n)$ the dynamics turns out to be sensitive to $n$. In particular, we observe the point particle $(n=0)$ scenario to be integrable whereas at higher $n$ the dynamics seems to be chaotic. }


\begin{document}

\maketitle
\flushbottom

\section{Introduction} 
The investigation of chaotic systems has captivated numerous prominent researchers across different fields for several decades. Defining chaos rigorously remains a subject of ambiguity to this day. However, it is now widely accepted that the sensitivity to the initial condition of a classical dynamical system subject to some constraints is believed to define a chaotic system. Here, we will also follow this definition accompanied by well-known chaotic indicators such as the Poincare section, Fast Lyapunov indicator(FLI), etc. 
%The relationship between the chaos and the (non)-integrability is another scorching issue in the context.

In a true sense, integrability means that a system of differential equations can be solved by a method of quadratures i.e. its solution can be obtained in a finite number of algebraic operations\cite{frolov_black_2017}. Most of the studies involve the existence of the integrals of motion for showing the integrability of a dynamical system. Due to the Liouville theorem, a Hamiltonian system of N degrees of freedom is said to be integrable if it has exactly N integrals of motion and these conserved quantities $Q_i= f(p,q)$ including the energy can be used to construct the solution where the corresponding phase space is 2N dimensional consisting of coordinates $q_i$ and the canonical momenta $p_i$. % Integrability gives the N conserved quantities $Q_i= f(p,q)$ including the energy.
These charges define a N-dimensional torus in the phase space. To every tori of rational winding number, there exists infinitesimally close to it, the tori having irrational winding number. Such a tori is known as KAM tori. According to the KAM theorem, when deformed the majority of the tori undergo a slight deformation but manage to survive. However, tori characterized by the rational frequency ratios experience destruction and therefore leading to chaotic motion on these tori. For such a system, integrability and chaos are considered to be complementary to each other.

The dynamical behavior of a point particle in various curved backgrounds has been extensively studied \cite{carter_global_1968,Bombelli:1991eg,Dettmann:1994dj,sota_chaos_1996,Chen:2002gz,page_complete_2007,hanan_chaotic_2007,chen_chaotic_2016}.
Around a black hole spacetime, the dynamics turn out to be integrable \cite{carter_global_1968}. 
%Equivalent investigations into the integrability of the system have been conducted on the $AdS$ side, by establishing the map between the integrable models. \cite{}. In view of the classical strings, finding the trajectory solution for the full phase space, by reducing to the quadratures is often not possible. 
However, the dynamics of such a geodesic becomes chaotic and non-integrable if we consider more complicated geometries like in \cite{hanan_chaotic_2007,chen_chaotic_2016,chervonyi_non-integrability_2014}. Moreover, the point particle dynamics near the horizon shows chaotic behaviour \cite{Dalui:2018qqv,hashimotoUniversalityChaosParticle2017,Dalui:2019esx}. 
Even after all these developments geodesics are not considered as the most suitable means for investigating the chaos produced by black holes. Despite the fact that geodesics being integrable in a large class of simple backgrounds, they exhibit thermalisation and chaos due to finite Hawking temperature. This motivates to work with the string dynamics. The works of Hadamard-Anosov and others \cite{basu_chaos_2011,basu_chaotic_2017,frolov_chaotic_1999,Basu:2011fw,Basu:2012ae,basuAnalyticNonintegrabilityString2011a} delving into the constant curvature spaces, have shed light on the sensitive instability of phase space trajectories within them- a characteristic feature of chaos. It appears that chaos is a recurring feature in the dynamics of semiclassical strings within AdS backgrounds \cite{zayas_chaos_2010, basu_chaos_2011,basuAnalyticNonintegrabilityString2011a,basuIntegrabilityLost2011} with the notable exceptions for e.g, $AdS_{5} \times S^{5}$ in which string dynamics
    turns out to be integrable, see \cite{beisertReviewAdSCFT2012a,vantongerenIntegrabilityAdSSuperstring2014,ricciTDualityIntegrabilityStrings2007,benaHiddenSymmetriesAdS2004}. All these advancements \cite{rigatosNonintegrabilityaC2020,ishii_fate_2017,basuIntegrabilityLost2011, basu_chaotic_2017,asanoChaoticStringsPenrose2015,hashimotoChaosWilsonLoop2018,Nunez:2018qcj,roychowdhuryAnalyticIntegrabilityStrings2017,banerjeeProbingAnalyticalNumerical2018a,maChaosRingString2014a,giataganasNonintegrabilityNonrelativisticTheories2014} collectively indicate that the class of string integrable backgrounds constitutes only a limited subset within the larger class of particle integrable backgrounds. The (non)-integrability aspect can also be considered when examining the behavior of string dynamics in Dp-branes \cite{stepanchuk_nonintegrability_2013,giataganasNonintegrabilityChaosUnquenched2017}. Previous studies \cite{stepanchuk_nonintegrability_2013} have shown that the classically extended string in Dp-brane in extremal scenarios having charge as the interpolation parameter between flat space and $AdS_5 \times S^5$ are non-integrable. In this paper, our aim is to explore this concept and conduct a numerical analysis of the circular string dynamics in a black p-brane. Specifically, we will focus on two cases: p=5 and p=6. We shall begin with the non-extremal scenarios first and then eventually examine the extremal situation as well by means of varying the suitable parameter.

Another interesting paradigm in recent times is the remarkable application of quantum information theoretic tools to study the quantum effects of gravity. Black holes are not only quantum in nature but also thermal and therefore share the basic property of chaos which in the light of quantum information are commonly postulated to be the fastest scramblers in Nature \cite{sekino_fast_2008}. For a finite temperature QFT, it is well-known that the defining characteristic of chaos is bounded by  what is called Maldacena, Shenker, Standford (MSS) bound given as:
$\lambda \le 2 \pi k_{B} T$ \cite{maldacena_bound_2016}. Currently, such a bound is saturated by holographic dual models such as SYK model \cite{sachdev_gapless_1993,marcus_new_2019,kitaev2015simple,kitaev2015simple2}. 
The relationship between the chaos bound and the study of string dynamics in p-black branes has not been explored thus far. A generalized bound inequality for string dynamics in AdS geometries has been predicted in \cite{cubrovic_bound_2019}. Our numerical investigation of the Lyapunov exponent associated with the orbits of the different string modes, to our surprise, suggests the existence of such a bound even though the bound was originally formulated for the field theories having classical gravity dual. Several efforts have been made to investigate the existence of such a bound in point particle dynamics\cite{Gao:2022ybw,gwak_violation_2022,hashimotoUniversalityChaosParticle2017,Kan:2021blg,leiCircularMotionCharged2022,daluiPresenceHorizonMakes2019,Hashimoto:2021afd}.%However, recently it has been noticed that the bound is violated even in point particle scenarios \cite{gwak_violation_2022,gao_chaos_2022,kanBoundLyapunovExponent2022}.

The rest of the paper is organized as follows. In section \ref{2} and \ref{sec: uncharged}, we analyze the pulsating string in non-extremal charged and neutral  p=5 brane respectively, using the underlying Hamiltonian dynamics. In the next subsection \ref{sec: p=6}, we repeat a similar analysis for p=6 brane. In section \ref{sec: winding}, we discuss the role of the winding number pertaining to our study. Finally, in section \ref{Discussion}, we summarize the numerical results of our analysis and give possible future directions.

\section{Black p-branes} \label{1}
 
 In this section, we shall briefly explain the black p-branes in D=10 dimensions. 
 %We will justify why we have taken specifically the p=5,6 branes. 
 %\newline 
% The black p-branes are solutions of vacuum Einstein equation $R_{\mu\nu}=0$ in higher dimensions and the corresponding metric is-
%\begin{equation}
%ds^{2} = -[1-(\frac{r_{0}}{r})^{7-p}]dt^{2}+[1-(\frac{r_{0}}{r})^{7-p}]^{-1} dr^2 +r^{2}d\Omega_{8-p}+\sum_{i=1}^{p} dx_{i}^{2}
%\end{equation}\label{eqn: brane_metric}
%Where $x_{1},x_{2},x_{3}...x_{p}$ represent spatial coordinates along the brane and $r_{0}$ represents event horizon radius. 
Non-extremal black p-branes are the solutions of 10-dimensional low energy string theory \cite{Horowitz:1991cd,becker2006string} and the corresponding metric ($p <7$) is:
\begin{equation}\label{eqn: brane_metric2}
ds^{2} = -\Delta_{+}\Delta_{-}^{-1/2}dt^{2}+ \Delta_{-}^{1/2}\sum_{i=1}^{p}dx_{i}^{2} + \Delta_{+}^{-1}\Delta_{-}^{\gamma}dr^{2}+r^{2} \Delta_{-}^{\gamma +1} d\Omega_{8-p}
\end{equation}
where $\Delta_{\pm} = 1-(\frac{r_{\pm}}{r})^{7-p}$, $\gamma = -\frac{1}{2} - \frac{5-p}{7-p}$ with
$r_{+}$ and $r_{-}$ representing outer horizon and inner horizon radii respectively. The charge and mass per unit p-volume of the black brane are respectively given by 
\vspace{3mm}
\begin{flalign*}
Q & = \frac{7-p}{2}(r_{+} r_{-})^{(7-p)/2}\\
M & = \frac{\Omega_{8-p}}{2k_{10}^2} \Big( (8-p)r_{+}^{7-p} - r_{-}^{7-p}  \Big)
\end{flalign*}
where $\Omega_{8-p}$ is the volume of unit (${8-p}$)- sphere and $k_{10}^2$ = $8\pi G_{10}$.
\newline

Black-p-branes are characterized by a non-zero Hawking temperature which can be derived by a series of transformations \cite{klebanov_entropy_1996,Ohshima:2005ha,duff_black_1996} as given below:
\vspace{3.5mm}
\begin{center}
 $r^{7-p} = \Tilde{r}^{7-p} + r_{-}^{7-p}$, \hspace{3mm}$\Tilde{r}_{+}^{7-p} = \mu ^{7-p} \cosh^{2}\beta $, \hspace{3mm} $\Tilde{r}_{-}^{7-p} = \mu ^{7-p} \sinh^{2}\beta$ %$H = 1+(\frac{r_{-}}{\Tilde{r}})^{7-p}$, \hspace{3mm} $f = 1 -(\frac{\mu}{\Tilde{r}})^{7-p}$.
 \end{center}
This reduces the metric \ref{eqn: brane_metric2} to the form:
\begin{equation}\label{eqn: abel}
 ds^{2} = H^{-1/2} \Big( -fdt^{2}+\sum_{i=1}^{p} dx_{i}^{2} \Big) + H^{1/2} \Big( f^{-1}dr^{2}+r^{2}d\Omega_{8-p} \Big)
 %\end{equation}
 \end{equation}
 where $H = 1 + (\frac{r_{-}}{r})^{7-p}$ and $f = 1-(\frac{\mu}{r})^{7-p}$
 \newline

  Then we need to expand the metric in the vicinity of horizon, followed by a Wick rotation of the time coordinate. The periodicity of the Euclidean time gives the inverse temperature \cite{Ohshima:2005ha, duff_black_1996}
 \begin{flalign*}
     \beta & = \frac{4\pi\mu \cosh\beta}{7-p}\\
    or \hspace{3mm}   \beta &  = 2 \pi \Big( \frac{2 r_{+} }{7-p} \Big[ 1- (\frac{r_{-}}{r_{+}})^{7-p} \Big ] ^{\frac{-5+p}{2(7-p)}}     \Big)
 \end{flalign*}
 \newline
 When $r_{-} = 0$, we have the neutral black brane solution:
\begin{equation}\label{eqn: brane_metric}
ds^{2} = -\Big( 1-(\frac{r_{0}}{r})^{7-p} \Big) dt^{2}+ \Big( 1-(\frac{r_{0}}{r})^{7-p} \Big) ^{-1} dr^2 +r^{2}d\Omega_{8-p}+\sum_{i=1}^{p} dx_{i}^{2}
\end{equation}

One can also obtain the supersymmetric extremal black brane solutions by setting $r_{+} = r_{-}$ and then performing a change of coordinate $\Tilde{r}^{7-p} = r^{7-p} - r_{+}^{7-p}$, we find %so that the new horizon is located at $\Tilde{r} = 0$ 
\begin{equation}
 ds^{2} = \Big( 1+(\frac{r_{0}}{r})^{7-p} \Big) ^{-1/2}\Big( -dt^{2}+\sum_{i=1}^{p} dx_{i}^{2}\Big) + \Big( 1+(\frac{r_{0}}{r})^{7-p} \Big) ^{1/2}\Big( dr^{2}+r^{2}d\Omega_{8-p} \Big)
 \end{equation}\label{eqn: brane_metric3}
 The integrability of strings and geodesics in such extremal backgrounds have been studied through Normal Variational Equations (NVE) approach in \cite{stepanchuk_nonintegrability_2013,chervonyi_non-integrability_2014}. Note that such an approach can be implemented via metric \ref{eqn: abel}. However
 %The rest of the paper is organized as follows. In section \ref{2} and \ref{sec: uncharged}, we analyze the pulsating string on  p=5 brane (charged and uncharged) using the underlying Hamiltonian dynamics. In the next subsection \ref{sec: p=6}, we will repeat a similar analysis for p=6 brane. Finally in section \ref{sec: winding}, we will discuss the role of the winding number pertaining to our study and comment more generically, on the relation with the bound of \ref[Maldecena,Shenker] $\lambda \le 2 \pi T $. 
%and as pointed out earlier the importance of the p=5,6 branes. Although the analysis we have done here, can be extended to other branes but we limit ourselves to the cases of p=5,6. Since we would like to probe the dynamics of  p-brane numerically therefore we shall focus particularly on two branes as pointed out earlier namely, p=5 and p=6 branes. The reason to consider these two branes apart from providing the matter of numerical convenience and 
our main objective is to study the behavior of 
 closed string in p-branes through numerical analysis. To achieve this goal, we will pivot ourselves particularly around two types of branes namely p=5 and p=6. These branes offer more feasible access in handling the situation numerically that can be realized. However, when dealing with branes of lower dimensions, we have to deal with more coordinates transverse to the brane which need to be eliminated. Typically, this is achieved through a consistent truncation method \cite{Cheung_2021, M_LLER_2006} but we shall completely ignore it and plan to address it in the future.

 \section{Circular string in p-brane}
 The propagation of a closed circular string in any arbitrary curved background can be modeled by the Polyakov action given as,
 \begin{equation} \label{eqn: polyakov}
 S = -\frac{1}{2\pi \alpha^{\prime}} \int d\sigma d\tau \sqrt{-g} g^{\alpha \beta} G_{\mu \nu}(\partial_{\alpha} X^{\mu}\partial_{\beta} X^{\nu})
 \end{equation}
 where $\alpha^{\prime} = l_s^{2}$ ($l_s$ represents the string length). $X^{\mu}$ represents the target space co-ordinates, $G_{\mu \nu}$ is the target space metric and $g_{\alpha \beta}$ is the worldsheet metric. We utilise the reparameterization and Weyl symmetries of the action and fix the conformal gauge, $g^{\alpha\beta}$ = $\eta^{\alpha\beta}$. In this gauge, the vanishing of energy-momentum tensor $T_{\alpha \beta}$ = 0 leads to the following constraints
 \begin{flalign}
 G_{\mu\nu} \partial_{\tau}X^{\mu} \partial_{\sigma}X^{\nu}& = 0 \label{eqn: gauge_conformal}\\
G_{\mu\nu} \Big( \partial_{\tau}X^{\mu} \partial_{\tau}X^{\nu}+\partial_{\sigma}X^{\mu} \partial_{\sigma}X^{\nu} \Big)& = 0\label{eqn: gauge_conformal2}
\end{flalign}
The target space metric $G_{\mu \nu}$ is given by equation \ref{eqn: brane_metric2}. Now, we use the pulsating string ansatz for p = 5 brane and the action \ref{eqn: polyakov}, to construct the Hamiltonian and the corresponding equation of motion.

%Now, we use the pulsating string ansatz and the action \ref{eqn: polyakov} to construct the Hamiltonian.  The corresponding equations of motion are highly nonlinear and coupled. This motivates us to look for  non-integrability/chaos of the string dynamics. 
%\newpage
\subsection{Black 5-brane}  \label{2}
We will consider the following ansatz representing the circular string 
\begin{eqnarray}
    t = t(\tau) \ ,  r= r(\tau) \ , \phi_{1} = \phi_{1}(\tau) \ , \phi_{2} = \phi_{2}(\tau) \ , \phi_{3} = n\sigma
\end{eqnarray}
where $n$ denotes the winding number of the string along $\phi_{3}$ direction. We assume the spatial coordinates $x^{i}$ are constant.\\
\newline
In this case, d$\Omega_{3}$ = $d\phi_{1}^{2} + \sin^{2}\phi_{1}d\phi_{3}^{2} + \cos^{2}\phi_{1}d\phi_{2}^{2}$. 
Substituting in the Polyakov action \ref{eqn: polyakov}, we get the following Lagrangian
\begin{multline*}
L = -\frac{1}{2\pi\alpha^{\prime}} \Big( \Delta_{+}\Delta_{-}^{-1/2} \dot t^{2}-\Delta_{+}^{-1}\Delta_{-}^{-1/2} \dot r^{2}-r^{2}\Delta_{-}^{1/2} ( \dot \phi_{1}^{2}+ \cos^{2}\phi_1 \dot \phi_2^{2})
+\Delta_{-}^{1/2}r^{2}n^{2} \sin^{2}\phi_{1} \Big)
\end{multline*}
where $\Delta_{\pm}(r) = 1-(\frac{r_{\pm}}{r})^2$ and dot represents the derivative with respect to $\tau$.\\

The corresponding Hamiltonian and the equations of motion can be obtained as follows:
\begin{multline*}
H = \frac{\pi \alpha^{\prime}}{2}\Big(\Delta_{+}\Delta_{-}^{1/2} p_{r}^{2}+\frac{p_{\phi_{1}^{2}}}{r^2 \Delta_{-}^{1/2}} +\frac{p_{\phi_{2}^{2}}}{r^2 \Delta_{-}^{1/2} \cos^{2}\phi_1} - \frac{p_{t}^{2}}{\Delta_{+}^{-1}\Delta_{-}^{1/2}} \Big) + \frac{1}{2 \pi \alpha^{\prime}}n^{2}r^{2}\Delta_{-}^{1/2} \sin^{2}\phi_{1}
\end{multline*}

\begin{flalign}
\dot{p_{t}} & = 0 \label{eqn: energy}\\
\dot{t} & = -\pi \alpha^{\prime}\Delta_{+}
\frac{p_{t}}{\Delta_{-}^{1/2}}\\
\dot{p_{r}} & = \frac{\pi \alpha^{\prime}}{2}\frac{\partial} {\partial r}\Big(p_{t}^{2} \Delta_{-}^{-1/2}\Delta_{+}-p_{r}^{2}
\Delta_{-}^{1/2}\Delta_{+} - (p_{\phi_{1}^{2}}+\frac{p_{\phi_{2}^{2}}}{\cos^{2}\phi_1})\frac{1}{r^2 \Delta_{-}^{1/2}}- \frac{n^{2}}{2\pi \alpha^{\prime}} r^2 \Delta_{-}^{1/2} \sin^{2}\phi_{1} \Big) \\  
\dot{r} & = \pi \alpha^{\prime} \Delta_{+}\Delta_{-}^{1/2} p_{r} \\
\dot{p}_{\phi_{1}} &  = \frac{-\pi \alpha^{\prime}}{r^{2}\phi_{1}}p_{\phi_{2}}^{2}\sec^{2}\phi_{1}\tan\phi_{1} - \frac{n^{2}}{\pi \alpha^{\prime}} r^{2}\Delta_{-}^{1/2} \sin\phi_{1} \cos\phi_{1}\\
\dot{\phi_{1}} & = \pi \alpha^{\prime} \frac{p_{\phi_{1}}}{r^2}\\
\dot{p}_{\phi_{2}} & = 0 \label{eqn: angular}\\
\dot{\phi_{2}} & = \pi \alpha^{\prime} \frac{p_{\phi_{1}}}{r^2 \Delta_{-}^{1/2} \cos^{2}\phi_{1}}\label{eqn: azimuthal}
\end{flalign}
Equations \ref{eqn: energy} and \ref{eqn: angular} give two constants of motion- $p_t$ = E (energy) and $p_{\phi_{2}}$ = l (angular momentum). The  constraint \ref{eqn: gauge_conformal} is trivially satisfied, however equation \ref{eqn: gauge_conformal2} gives H = 0.
%\subsubsection{Numerical Analysis}
\newline
\subsubsection*{A. String trajectory for different values of charge}
\begin{figure}[ht!]
\begin{subfigure}
  \centering
  \includegraphics[width=0.492\textwidth]{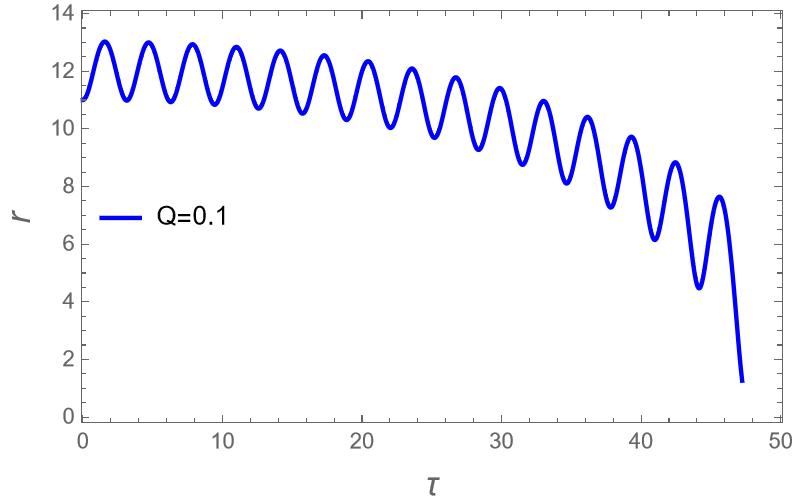}
 \put(-210,135){(a) \hspace{2cm} r(0)=11}
 % \phantomsubcaption\label{fig:nuvr}
\end{subfigure}
\hspace{2.8mm}
\begin{subfigure}
  \centering
  \includegraphics[width=0.492\textwidth]{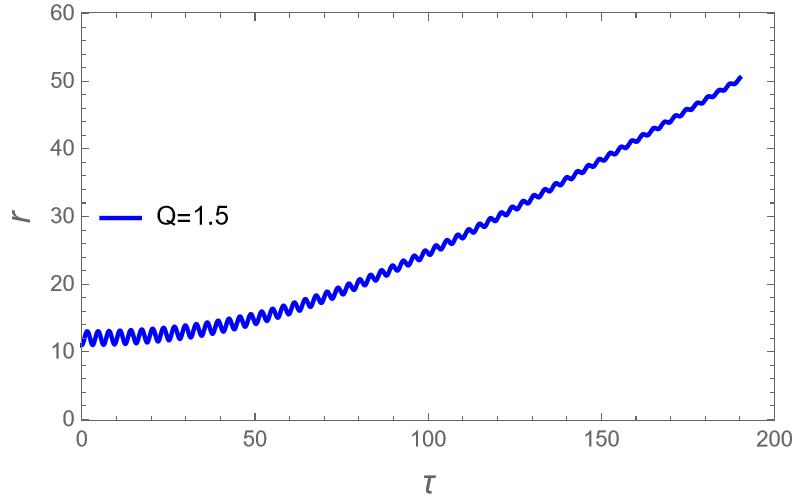}
\put(-210,135){(b) \hspace{2cm} r(0)=11}
%\put(-160,105){{\tiny r(0)=11}}
  %\phantomsubcaption\label{fig:varnuvsr}
\end{subfigure}
\vspace{2mm}
\begin{subfigure}
  \centering
  \includegraphics[width=0.494\textwidth]{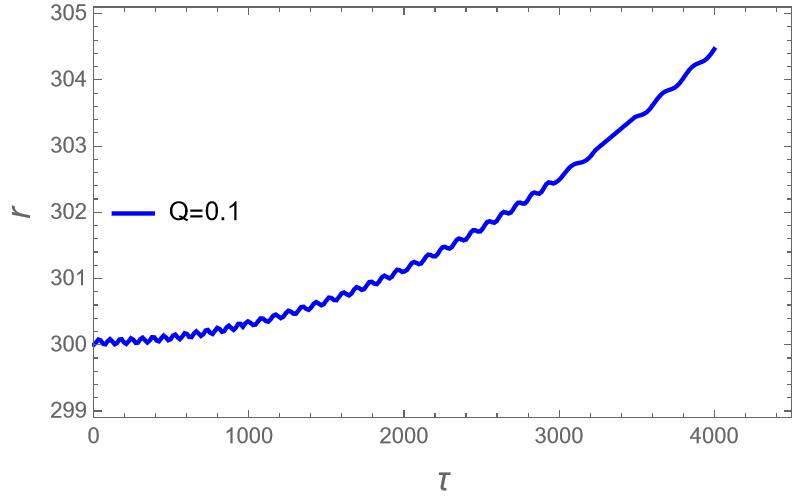} 
\put(-208,135){(c) \hspace{2cm} r(0) = 300}
  %\phantomsubcaption\label{fig:ensvr}
\end{subfigure}
\hspace{2.8mm}
\begin{subfigure}
  \centering
  \includegraphics[width=0.499\textwidth]{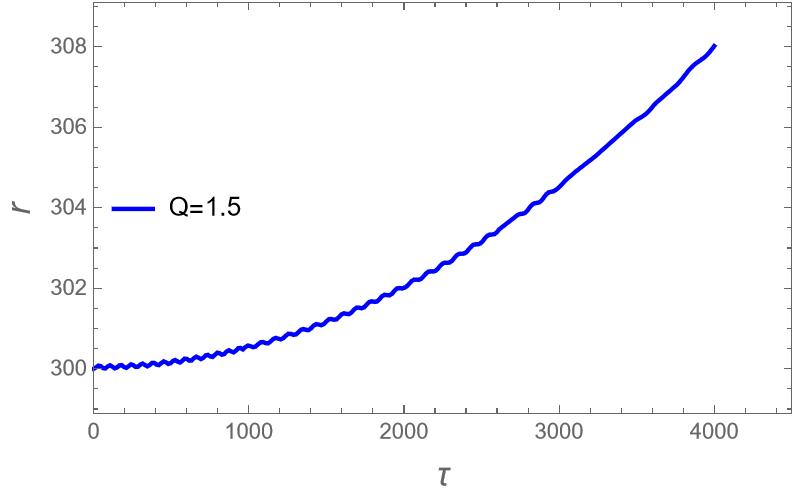}
\put(-213,138){(d) \hspace{2cm} r(0) = 300}
 % \phantomsubcaption\label{fig:varensvr}
\end{subfigure}
\caption{Plot showing the evolution of radial coordinate for different charges (Q=0.1,1.5) with different initial conditions r(0) = 11 (top panel) and r(0) = 300 (bottom panel).}\label{fig:orbit_p=5}
\end{figure}

Without loss of generality, we set the following initial conditions and parameters-
\hspace{1.0mm} $p_{r}(0) = 0$, $\phi_1(0)$ = 0, E = 7, l = 8  to numerically solve the equations of motion and we set $\pi \alpha^{\prime}$ = 1, $n$ = 1, $G_{10} = c = 1$ throughout the rest of the paper. We fix the mass M = $3\pi/8$, then  the charge  and 
 mass satisfy the inequality $Q \le 4 M/\pi$ for unit volume.  Considering the charge Q as a control parameter, we monitor the string dynamics by varying Q. For the visualisation of string dynamics, we mostly follow the procedures of \cite{ma_chaotic_2020}.  In figure \ref{fig:orbit_p=5}, we present the string trajectory for two different initial radial positions of the string.
 First, we consider the string initially close to the  brane (figure \ref{fig:orbit_p=5}(a),(b)). For a small charge (Q = 0.1), we find that  the string gets trapped in the black brane very quickly, whereas in the extremal limit (Q = 1.5) the string deviates from the horizon and rapidly escapes to infinity. However, when we keep the string at a large distance away from the  brane (figure \ref{fig:orbit_p=5}(c),(d)), the string escapes to infinity for both small and large value of charge. However, the rate of escape is very small unlike the first screnario (figure \ref{fig:orbit_p=5}(b)).
\newline
\subsubsection*{B. Fast Lyapunov Indicator}
\begin{figure}[ht!]
\begin{subfigure}
  \centering
  \includegraphics[width=0.48\textwidth]{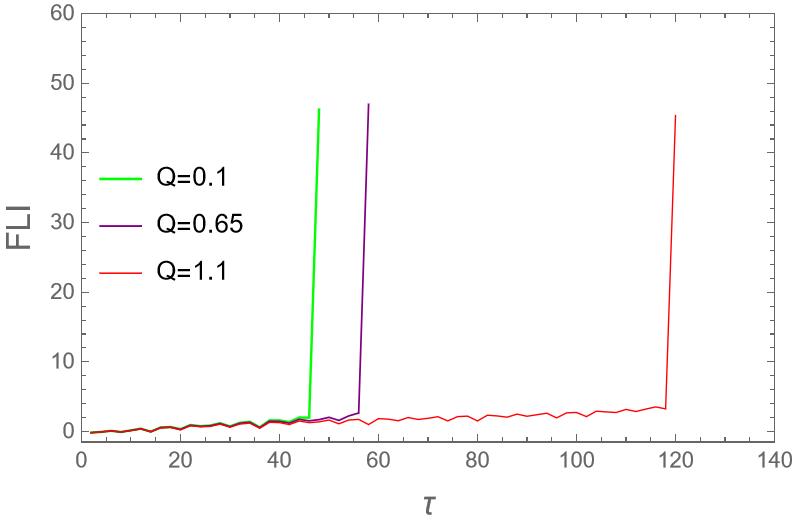}
\put(-210,135){(a) \hspace{2cm} r(0)=11}
 % \phantomsubcaption\label{fig:nuvr}
\end{subfigure}
%\hspace{2.6mm}
\begin{subfigure}
  \centering
  \includegraphics[width=0.48\textwidth]{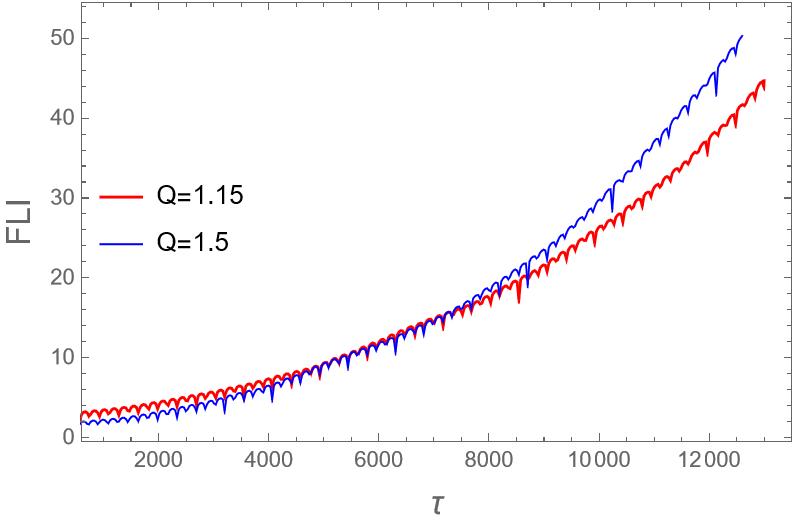}
\put(-210,138){(b) \hspace{2cm} r(0)=11}
  %\phantomsubcaption\label{fig:varnuvsr}
\end{subfigure}
%\vspace{2mm}
\begin{subfigure}
  \centering
  \includegraphics[width=0.48\textwidth]{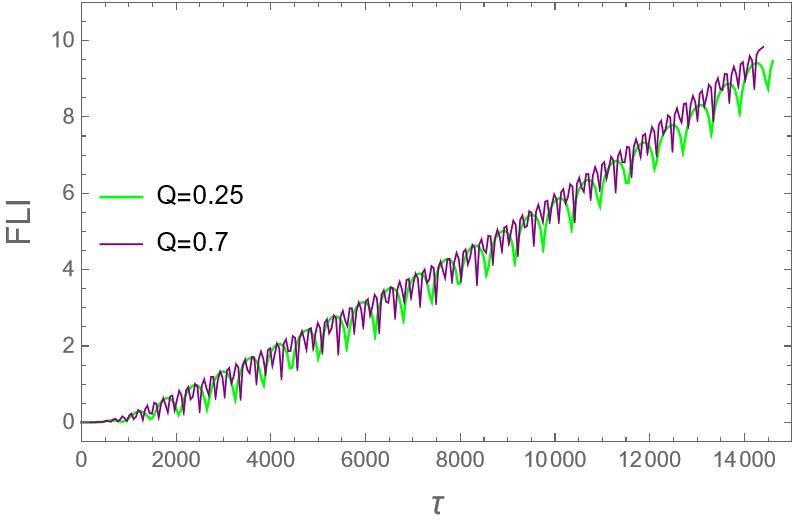} 
\put(-210,138){(c) \hspace{2cm} r(0)=300}
  %\phantomsubcaption\label{fig:ensvr}
\end{subfigure}
\hspace{2.8mm}
\begin{subfigure}
  \centering
  \includegraphics[width=0.49\textwidth]{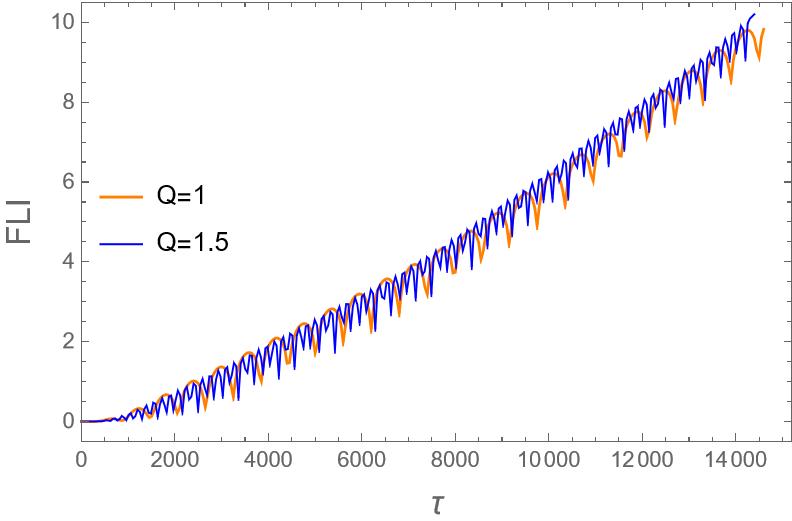}
\put(-210,140){(d) \hspace{2cm} r(0)=300}
 % \phantomsubcaption\label{fig:varensvr}
\end{subfigure}
\caption{Plot showing the behaviour of FLI for different Q with different initial conditions r(0) = 11 (top panel) and r(0) = 300 (bottom panel).}\label{fig: FLI_P=5}
\end{figure}
One of the salient features of the chaotic system is that its dynamics are extremely sensitive to the initial condition for which one has to study the chaotic indicators. 
The Largest Lyapunov Exponent(LLE) is a commonly used method which is based on the algorithm of measuring the average separation between the two initially nearby trajectories to characterize the nature (regular or chaotic) of orbits. However, there are some subtilities regarding LLE, for instance, sometimes it costs a large computation time to achieve a stable limiting value. Also, LLE is not suitable for distinguishing different regular orbits. Reference \cite{PhysRevD.74.083001} shows that the value of LLE is not co-ordinate invariant and so not reliable for relativistic systems. Some of these issues can be overcome by implementing a closely related method known as Fast Lyapunov Indicator (FLI). FLI is quicker for detecting chaos and order and easier to implement.  A detailed discussion on the Lyapunov indicators can be found in \cite{froeschle1997fast,article,ma2016application}.
In this work, we mostly employ FLI \cite{ma_chaotic_2020,wu_computation_2003,froeschle1997fast} which is based on the following  definition: %for two nearby trajectories separated initially by $d(0)$ and having separation $d(t)$ at longer time $t$ we define:
\begin{equation}
    \textit{FLI}= \log_{10}\frac{||\textbf{d}(t)||}{|| \textbf{d}(0)||}=  \lambda  t
\end{equation}
where $\lambda$ is the LLE, \textbf{d}(0) represents the initial separation between two 
 nearby trajectories and \textbf{d}(t) represents the separation at time t.
Thus from this equation, the FLI increases linearly (approximately) with $t$ for chaotic orbit with a positive slope whereas the slope is equal to zero for integrable motion. Note that, if the slope is very close to zero then orbit is quasi-periodic. In other words, the distance between two orbits increases exponentially for chaotic orbit and linearly (approximately) for non-chaotic orbit. %The slope of the FLI vs $\tau$ gives the $\lambda$ and therefore, if the slope is zero then integrable, if slope is strictly greater than zero then the orbit is chaotic and finally if the slope is near to zero then orbit is quasi-periodic. 
For practical computation, we use the following expression:
\begin{equation}
    FLI(t)= -k(1+ \log_{10} || d(0) ||) + \log_{10}\frac{||d(t)||}{|| d(0)||}
\end{equation}
where $k (k=0,1,2...)$ denotes the number of renormalization. We choose $||d(0)|| \approx 10^{-7}$ and $||d(t)|| = 0.1$ as critical value to implement the process of renormalization.

Now, in order to speculate the string dynamics, especially on the intermediate charges, we present FLIs in figure \ref{fig: FLI_P=5}. When $r(0) = 11$ and Q = 0.1, 0.65, 1.1 (figure \ref{fig: FLI_P=5}(a)), the FLI curve increases linearly at the beginning and then becomes vertical. The underlying reason for the sudden jump is due to the capture of the string in the black brane. Because of this collapse of string orbit, the calculation stopped at that instance and with the further increase of that critical times, we observe a shoot-up in  FLI. Note that the time of capture increases with charge. However, when $Q > 1.1$ (figure \ref{fig: FLI_P=5}(b)), the FLI increases approximately linearly and the string escapes to infinity. The rate of escape increases very slightly with charge as evident from the corresponding value of FLI.  Next, we concentrate on $r(0) = 300$ (Figure \ref{fig: FLI_P=5}(c),(d)). Here, the corresponding FLI curve shows a linear growth and the string escapes to infinity with uniform rate for all values of charges.  We clearly observe the consistencies between Figure \ref{fig:orbit_p=5} and \ref{fig: FLI_P=5}. Thus by tuning control parameter(charge), we can visualise different chaotic modes.
 \vspace{3.5mm}
\subsection{Neutral black 5-brane}\label{sec: uncharged}
\begin{figure}[!ht]
 \begin{subfigure}%%{0.485\textwidth}
 \centering
  \includegraphics[width=0.485\textwidth]{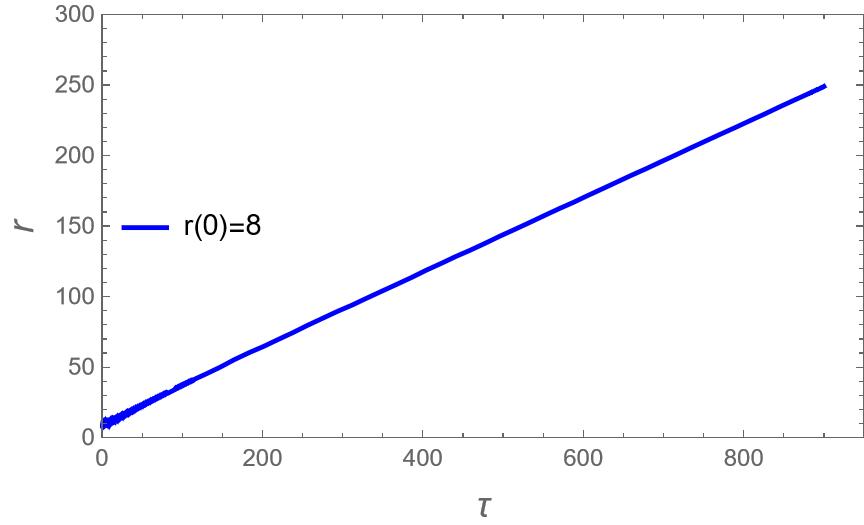}
  \put(-210,130){(a)}
  %\phantomsubcaption\label{fig:varnuvsr}
\end{subfigure}
\hfill
\begin{subfigure}%[b]%{0.47\textwidth}
\centering
\includegraphics[width=0.47\textwidth]{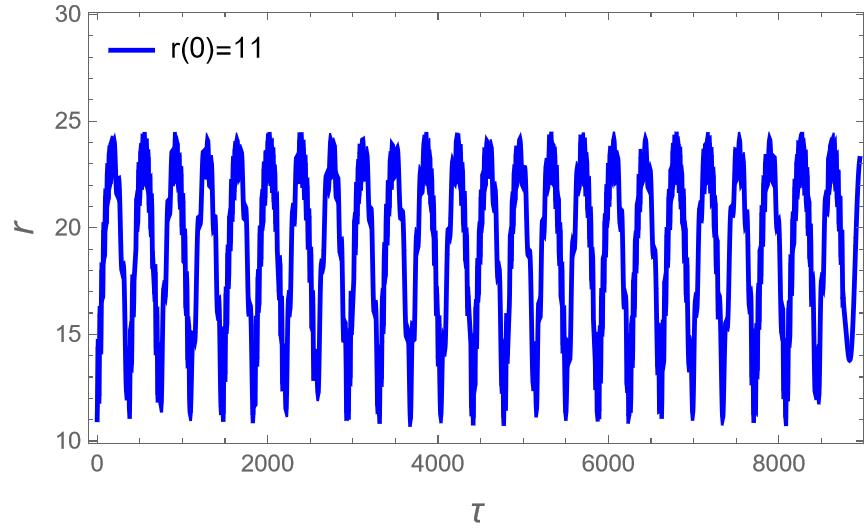}
\put(-200,130){(b)}
\caption{Plot showing the time evolution $r(\tau)$ indicating (a) escape to infinity and (b) quasiperiodic behavior of the string. For both the plots, we use E = 9, l = 8, n = 1,  $r_{+}$ = 1. The initial conditions are $p_{r}(0)$ = 0, $\phi_{1}(0)$ = 0, r(0) = 8 for (a) and $p_{r}(0)$ = 0, $\phi_{1}(0)$ = 0, r(0) = 11 for (b).}\label{fig: uncharged radial p=5}
\end{subfigure}
\end{figure}

Now, we investigate the chaotic dynamics by setting Q = 0. Then the metric \ref{eqn: brane_metric} reduces to the product of  5-dimensional Euclidean space and 5-dimensional Schwarzschild one:
\begin{equation*}\label{eqn: neutral}
ds^{2} = -\Big( 1-(\frac{r_{0}}{r})^{2} \Big) dt^{2}+ \Big( 1-(\frac{r_{0}}{r})^{2} \Big) ^{-1} dr^2 +r^{2}d\Omega_{3}+\sum_{i=1}^{5} dx_{i}^{2}
\end{equation*}
With the same ansatz as we assumed in the beginning, we find the Hamiltonian 
\begin{equation}
H = \frac{\pi \alpha^{\prime}}{2} \Big( \Delta_{+} p_{r}^{2}+\frac{p_{\phi_{1}^{2}}}{r^2}+\frac{p_{\phi_{2}^{2}}}{r^{2} \cos^{2}\phi_{1}} - \frac{p_{t}^{2}}{\Delta_{+}} \Big) + \frac{1}{2 \pi \alpha^{\prime}}n^{2}r^{2} \sin^{2}\phi_{1}
\end{equation}
 The equations of motion are \ref{eqn: energy}-\ref{eqn: azimuthal} with $\Delta_{-} = 1$. After solving the equations of motion, one obtains various possible modes of the string. Here, we present the two asymptotic modes of the string - escape to infinity and long times oscillations around the event horizon in figure \ref{fig: uncharged radial p=5}. We comment on the chaotic behaviour  by numerically evaluating the Poincare  section and Fast Lyapunov Indicator. %However, we would like to point out that  the quasiperiodic behavior  of an observable is not necessarily non-chaotic as the time
 %series possibly involve incommensurate  frequencies. We clarify this behavior by numerically evaluating  the Poincare  section and Fast Lyapunov Indicator.
 \vspace{5mm}
\subsubsection*{A. String trajectory}
\newpage
\subsubsection*{B. Poincare section}
\begin{figure}[!ht]
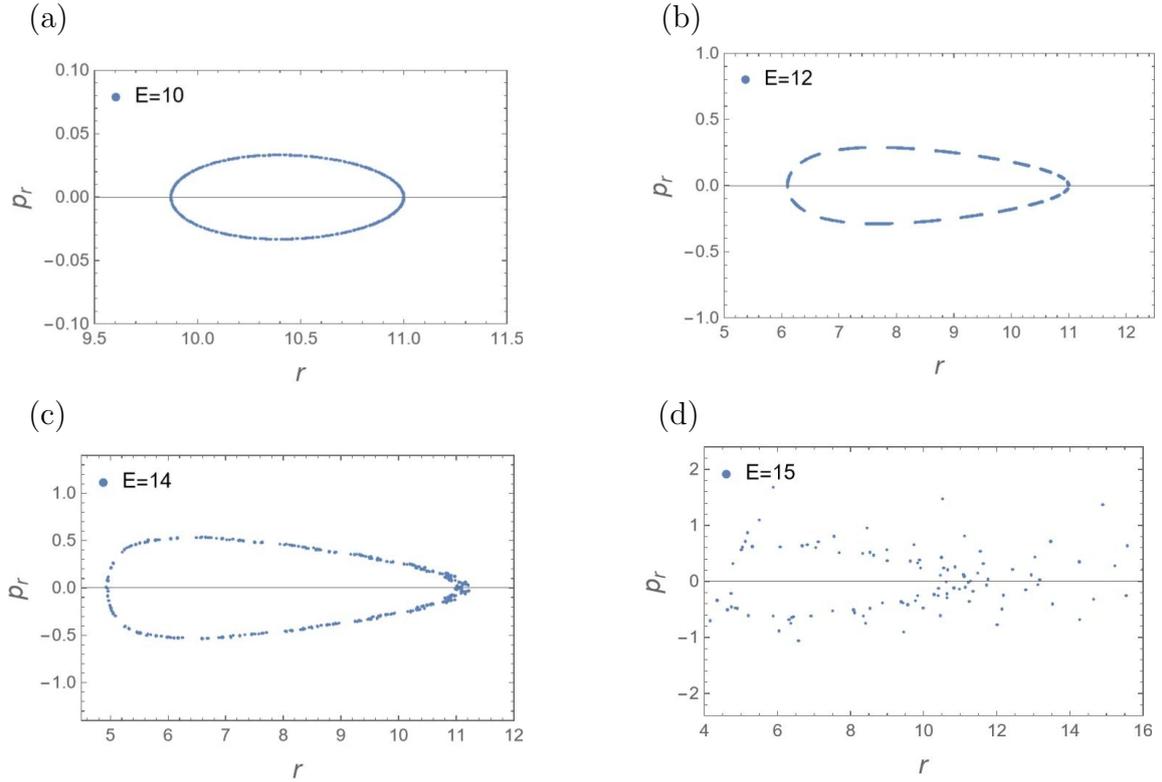

\begin{subfigure}
  \centering
  \includegraphics[width=0.45\textwidth]{poincareE=10.pdf}
 \put(-185,135){(a)}
 % \phantomsubcaption\label{fig:nu
 %vr}
\end{subfigure}
%\hspace{2.6mm}
\begin{subfigure}
  \centering
  \includegraphics[width=0.44\textwidth]{poincareE=12.pdf}
\put(-185,135){(b)}
  %\phantomsubcaption\label{fig:varnuvsr}
\end{subfigure}
\begin{subfigure}
  \centering
  \includegraphics[width=0.45\textwidth]{poincareE=14.pdf} 
\put(-185,135){(c)}
  %\phantomsubcaption\label{fig:ensvr}
\end{subfigure}
\hspace{12mm}
\begin{subfigure}
  \centering
  \includegraphics[width=0.45\textwidth]{poincareE=15.pdf}
\put(-185,135){(d)}
 % \phantomsubcaption\label{fig:varensvr}
\end{subfigure}
\caption{Plot showing the nature of poincare sections on the plane $\phi_1$   = 0 for (a) E = 10  (b) E = 12 (c) E = 14 (d) E = 15 indicating the distortion of tori with the increase of energy. 
We have set $r(0) = 11$, $p_{r}(0)$ = 0, $\phi_{1}(0)$ = 0, l = 8, n = 1, $r_{+}$ = 1.}\label{fig: poincare_section}
\end{figure}

%A Hamiltonian system of N degrees of freedom is said to be integrable if it has exactly N constants of motion. Because of the presence of large conservation laws, their dynamics are constrained in the phase and confined to invariant tori known as KAM tori defined by N constants of motion. But if the system is weakly perturbed by a integrability breaking parameter($\epsilon$), then some of these tori are distorted. With the increase of $\epsilon$, most of the tori get destroyed and eventually the phase space comprises of discrete points. At this stage, majority of the phase space is accessible to the system and the system becomes chaotic.\\
 Taking energy E as a control parameter, we provide the Poincare sections in the phase space (r, $p_r$) corresponding to different energies in figure \ref{fig: poincare_section}. From the first two plots, we observe a  quasi-periodic nature of the KAM tori  when E = 10 and E = 12. However, the tori starts deforming with the further increase of energy ($E > 12$) and finally, when E = 15, we see a complete deformation and a collection of discrete points  of the tori. Thus, with the increase of energy, our system becomes more and more chaotic.
 \newline
 
\subsubsection*{C. Fast Lyapunov Indicator}
\begin{figure}
     \centering
     \includegraphics[width=0.7\textwidth]{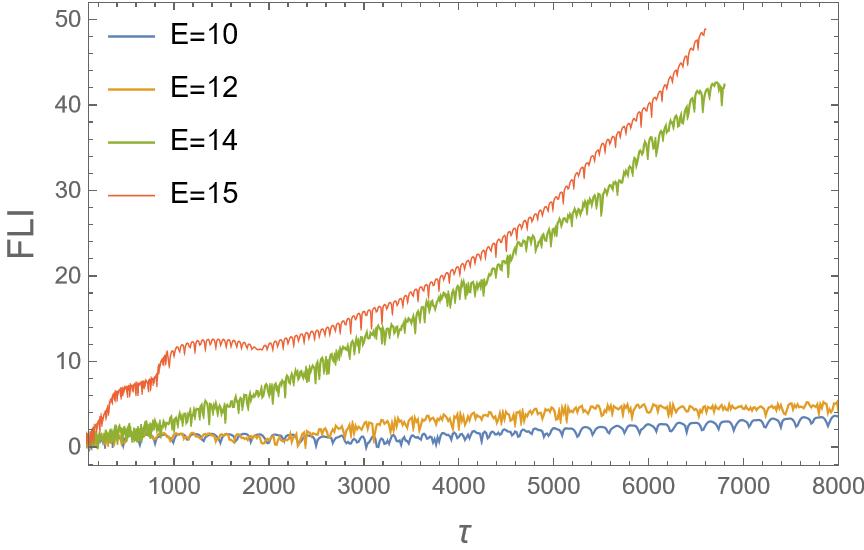}
%\put(-185,135){r(0)=11}
     \caption{Plot showing the behaviour of FLI for $r(0) = 11$ in uncharged p=5 brane. The other parameters are same as that of figure \ref{fig: poincare_section}.}
      \label{fig: FLI(uncharged)}
     \end{figure}

We also present FLI in figure \ref{fig: FLI(uncharged)} for the corresponding energies. Note that for E=10 and E = 12, the corresponding string orbit is quasiperiodic (characterised by almost zero slope). However, with the increase of energy, the FLI increases almost linearly with a much higher slope. Thus, transition to chaos with the increase of E is consistent with figure \ref{fig: poincare_section}.%In the left figure (E=10,12), we observe a persistent oscillation in the curve indicating a quasiperiodic nature of the orbit  whereas the right figure (E=14,15) shows an exponential growth of separation with time. Thus, we observe a transition from quasiperiodic to chaos with the increase of energy consistent with figure \ref{fig: poincare_section}.
 \newpage

\begin{figure}[!ht]
    \centering
    \includegraphics[width = 0.7\textwidth]{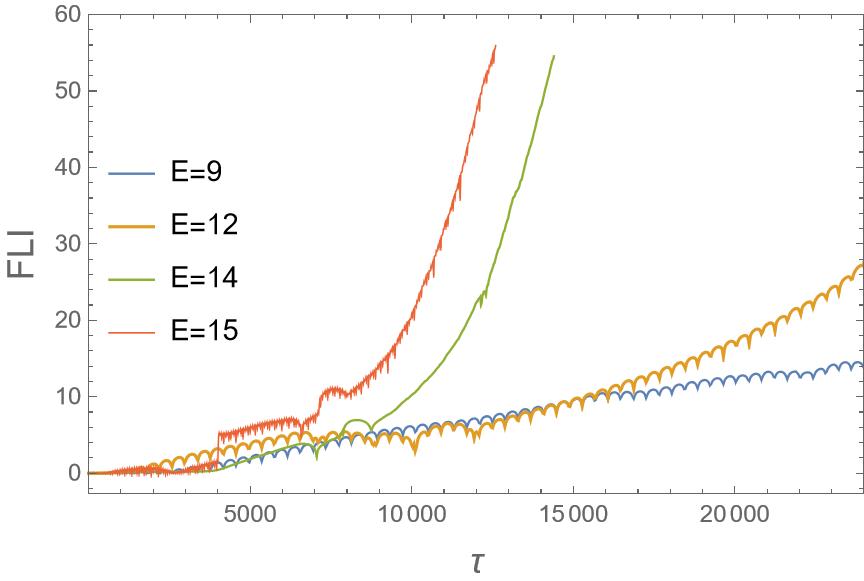}
    %\put(-200,165){r(0)=300}
    \caption{Plot showing the behaviour of FLI for $r(0) = 300$ in uncharged p=5 brane. The other parameters are same as that of figure \ref{fig: poincare_section}.}
    \label{fig:FLI(uncharged)2}
\end{figure}

%\begin{figure}[!ht]
 %\begin{subfigure}
 %\centering
 % \includegraphics[width=0.495\textwidth]{FLIP=5_E=10,12.pdf}
 % \put(-210,140){(a)}
%\end{subfigure}
%\begin{subfigure}
%\centering
%\includegraphics[width=0.49\textwidth]{FLIP=5_E=14,15_r=11.pdf}
%\put(-210,140){(b)}
%\end{subfigure}
%\caption{Plot showing the behaviour of FLI for $r(0) = 11$ in uncharged p=5 brane. The other parameters are same as that of figure \ref{fig: poincare_section}.}\label{fig: FLI(uncharged)}
%\end{figure}
\vspace{3.5mm}
As we increase the initial radial coordinate of the string (r(0)=300), we find a similar transition i.e the disappearance of the quasiperiodic behaviour and transition to more and more chaotic behaviour with the increase of energy (figure \ref{fig:FLI(uncharged)2}). %tendency of the string  escaping to infinity with the increase of energy (figure \ref{fig:FLI(uncharged)2}). %However, the FLI curve shows that there is no quasiperiodic orbit even at E = 12.
\vspace{3.5mm}
\newline
We now focus on the behaviour of pulsating string in p = 6 brane. In the next subsection, we construct our Hamiltonian and the equations of motion following the string ansatz and continue our discussion on chaotic dynamics.

\newpage
\subsection{Black 6-brane}\label{sec: p=6}
We make the following ansatz for the  circular string
\begin{center}
$t= t(\tau)$,\hspace{3mm} $r= r(\tau)$,\hspace{3mm} $\phi_{1} = \phi_{1}(\tau)$,\hspace{3mm}  $\phi_{2} = n\sigma$
\end{center}
where $n$ represents the winding number of the string along $\phi_{2}$ direction.
\newline
In this case, d$\Omega_{3}$ = $d\phi_{1}^{2} + sin^{2}\phi_{1}d\phi_{2}^{2} $. Substituting in the Polyakov action  equation \ref{eqn: gauge_conformal} we get the following Lagrangian 
\begin{equation*}
L = -\frac{1}{2\pi\alpha^{\prime}} \Big(\Delta_{+}\Delta_{-}^{-1/2}\dot{t}^{2}-\Delta_{+}^{-1}\Delta_{-}^{1/2}\dot{r}^{2}-r^{2}\Delta_{-}^{3/2}\dot{\phi_{1}}^{2}+\Delta_{-}^{3/2}r^{2}n^{2}\sin^{2}\phi_{1} \Big)
\end{equation*}
\newline
The corresponding Hamiltonian and the equation of motion are as follows:
\begin{equation*}
H = \frac{\pi \alpha^{\prime}}{2} \Big( \Delta_{+}\Delta_{-}^{-1/2} p_{r}^{2}+\frac{p_{\phi_{1}^{2}}}{r^2 \Delta_{-}^{3/2} } - \frac{p_{t}^{2}}{\Delta_{+}\Delta_{-}^{-1/2}} \Big) + \frac{1}{2 \pi \alpha^{\prime}}n^{2}\Delta_{-}^{3/2}r^{2}\sin^{2}\phi_{1}
\end{equation*}
\begin{flalign}
\dot{p_{t}} & = 0 \label{eqn: energy2}\\
\dot{t} & = -\pi \alpha^{\prime}\Delta_{-}^{1/2}
\Delta_{+}^{-1} p_{t}\\
\dot{p_{r}} & = \frac{\pi \alpha^{\prime}}{2}\frac{\partial} {\partial r} \Big( -p_{r}^{2}\Delta_{-}^{-1/2}\Delta_{+}+p_{t}^{2}
\Delta_{-}^{1/2}\Delta_{+}^{-1} - p_{\phi_{1}^{2}} \frac{1}{r^2 \Delta_{-}^{3/2}}- \frac{n^{2}}{2\pi \alpha^{\prime}} r^2 \Delta_{-}^{3/2} \sin^{2}\phi_{1} \Big)\\  
\dot{r} & = \pi \alpha^{\prime} \Delta_{+}\Delta_{-}^{-1/2} p_{r} \\
\dot{p}_{\phi_{1}} &  = - \frac{n^{2}}{\pi \alpha^{\prime}} r^{2}\Delta_{-}^{3/2} \sin\phi_{1}\cos\phi_{1}\\
\dot{\phi_{1}} & = \pi \alpha^{\prime} \frac{p_{\phi_{1}}}{r^2 \Delta_{-}^{3/2}}  
\end{flalign}
\newline
The conformal gauge constraint gives  H = 0 and the only constant of motion is the energy ($p_{t} = E)$.\\
\newline
%\subsubsection{Numerical Analysis}
\subsubsection*{A. String trajectory for different values of charge}
\begin{figure}[!ht]
\begin{subfigure}
  \centering
  \includegraphics[width=0.485\textwidth]{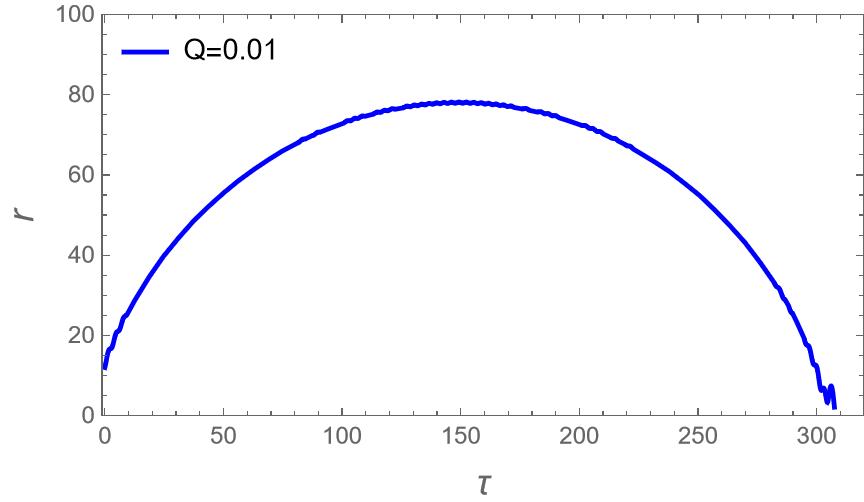}
\put(-202,130){(a) \hspace{2cm} r(0)=12}
 % \phantomsubcaption\label{fig:nuvr}
\end{subfigure}
\hspace{1.6mm}
\begin{subfigure}
  \centering
  \includegraphics[width=0.482\textwidth]{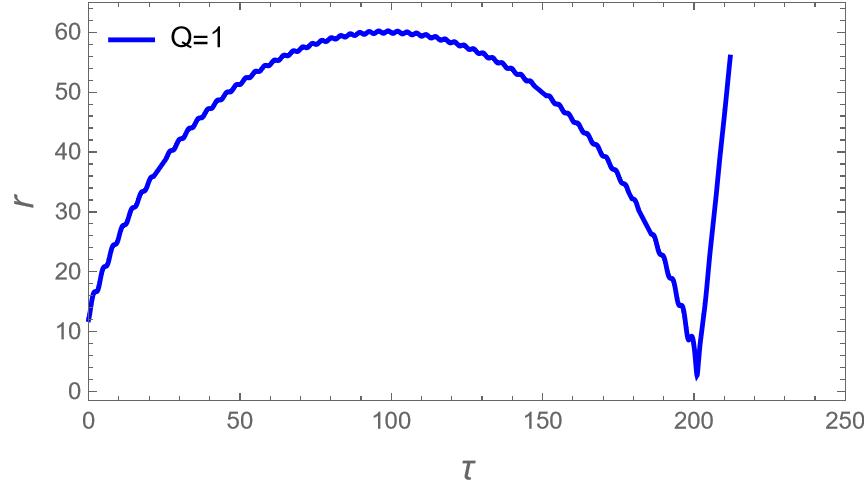}
 \put(-200,130){(b) \hspace{2cm} r(0)=12}
  %\phantomsubcaption\label{fig:varnuvsr}
\end{subfigure}
%\vspace{2mm}
\begin{subfigure}
  \centering
  \includegraphics[width=0.482\textwidth]{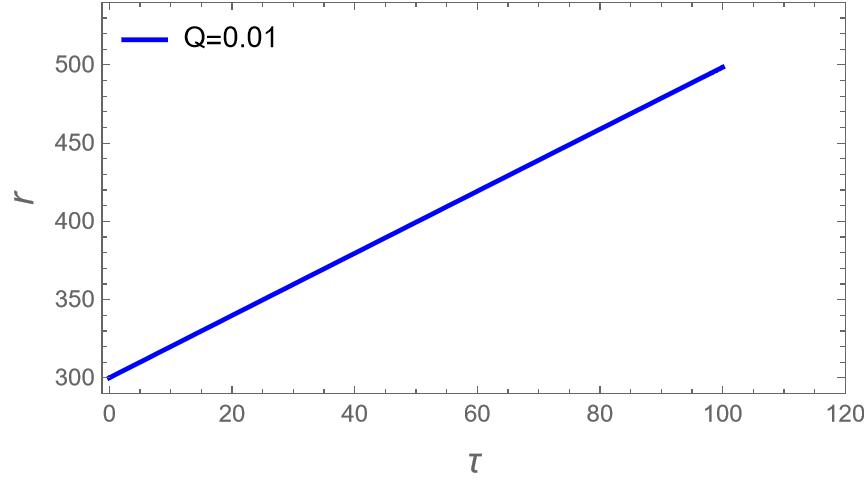} 
 \put(-202,125){(c) \hspace{2cm} r(0)=300}
  %\phantomsubcaption\label{fig:ensvr}
\end{subfigure}
\hspace{2.8mm}
\begin{subfigure}
  \centering
  \includegraphics[width=0.485\textwidth]{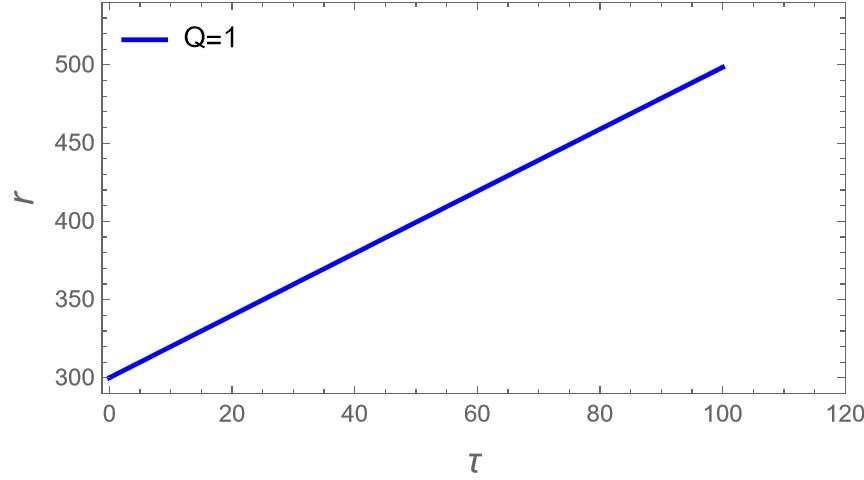}
\put(-200,125){(d) \hspace{2cm} r(0)=300}
 % \phantomsubcaption\label{fig:varensvr}
\end{subfigure}
\caption{Plot showing the evolution of radial coordinate for different charges (Q=0.01,1) with different initial conditions r(0) = 12 (top panel) and r(0) = 300 (bottom panel).}\label{fig: orbit_p=6}
\end{figure}

Once again, we study the dynamics by increasing the
charge Q for two different initial positions of the string. Without loss of generality, our choice of initial conditions and parameters are -
\hspace{1.5mm}
 $p_{r}(0) = 2$, $\phi(0)$ = 0, E = 7. We  also set the mass M = 0.5, then the charge  and 
 mass satisfy the inequality $Q \le 2 M$. When we consider the string initially  close to the brane, figure \ref{fig: orbit_p=6}(a),(b) reflects the  capture mode in the small charge limit and escape  mode in the extremal limit. However, only the escape mode survives when the string is initially at a large distance away from the brane (figure \ref{fig: orbit_p=6}(c),(d)). In the latter case, the overall dynamics is not much distinct from the corresponding p=5 case. To show this, we numerically evaluate FLI and display for different charges in figure \ref{fig: FLI_p=6}.   \\
\vspace{3mm}
%\vspace{3mm}
\subsubsection*{B. Fast Lyapunov Indicator}
\begin{figure}[!ht]
\begin{subfigure}
  \centering
  \includegraphics[width=0.48\textwidth]{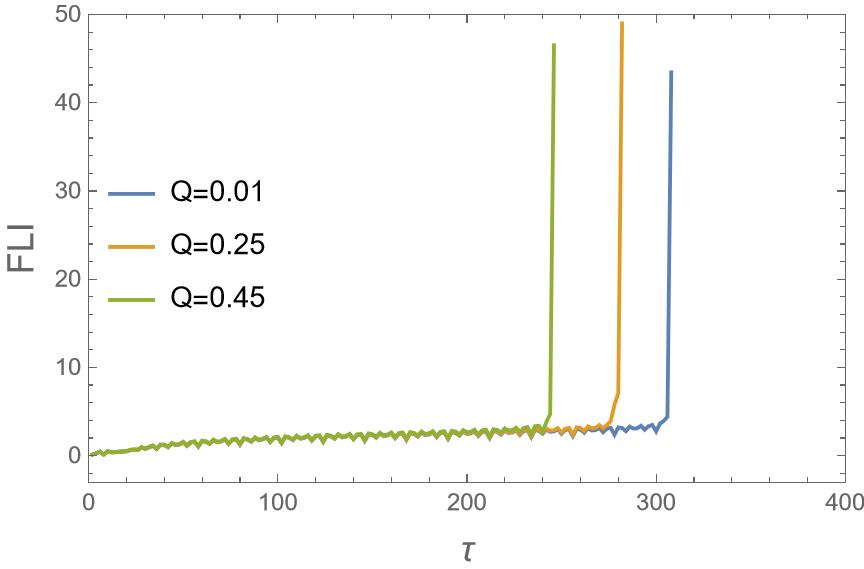}
 \put(-210,138){(a) \hspace{2cm} r(0)=12}
 % \phantomsubcaption\label{fig:nuvr}
\end{subfigure}
%\hspace{2.6mm}
\begin{subfigure}
  \centering
  \includegraphics[width=0.48\textwidth]{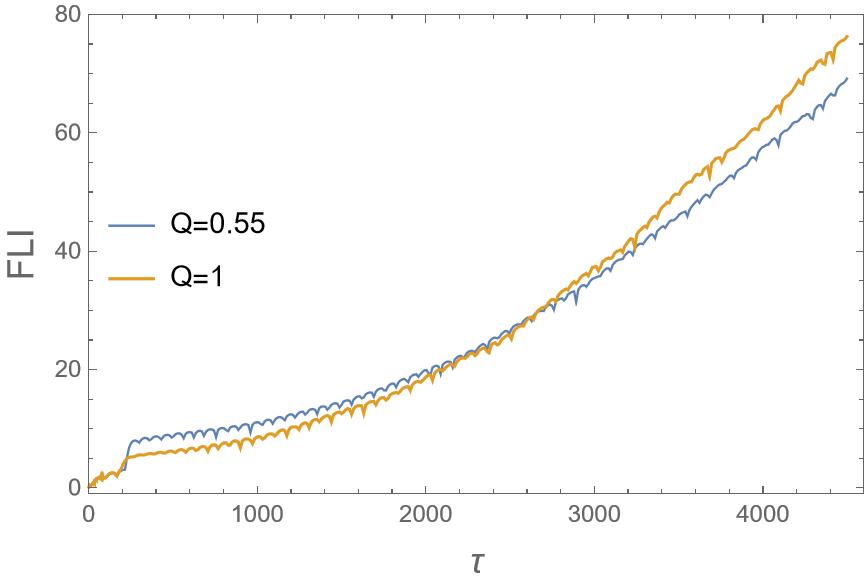}
\put(-210,138){(b) \hspace{2cm} r(0)=12}
  %\phantomsubcaption\label{fig:varnuvsr}
\end{subfigure}
%\vspace{2mm}
\begin{subfigure}
  \centering
  \includegraphics[width=0.48\textwidth]{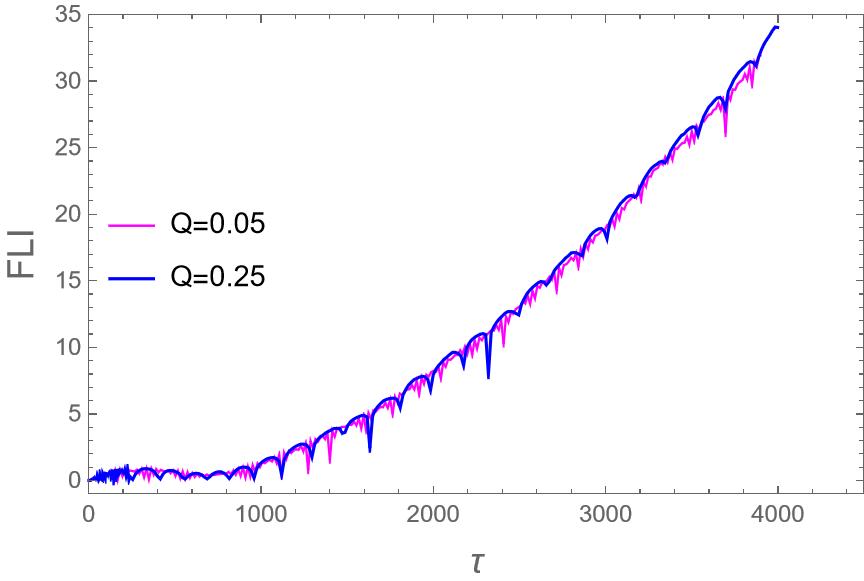} 
 \put(-210,140){(c) \hspace{2cm} r(0)=300}
  %\phantomsubcaption\label{fig:ensvr}
\end{subfigure}
\hspace{2.8mm}
\begin{subfigure}
  \centering
  \includegraphics[width=0.482\textwidth]{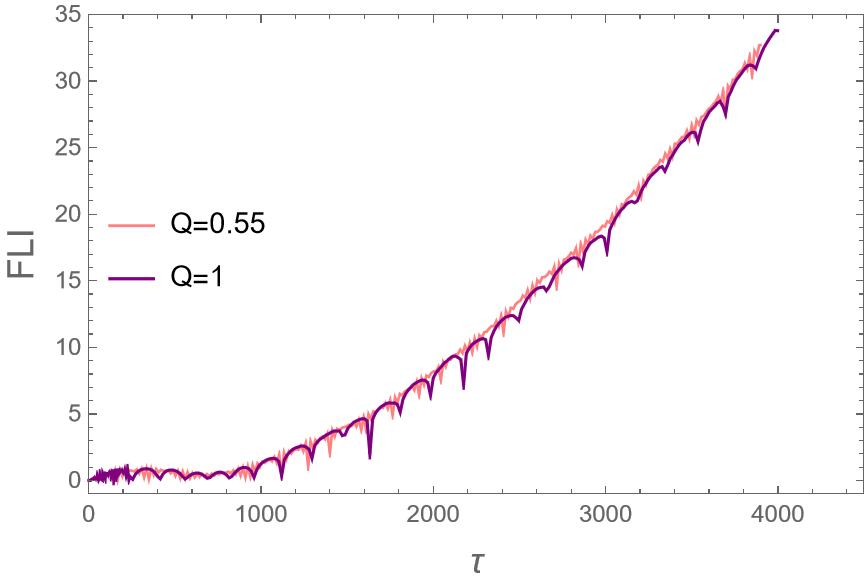}
\put(-210,140){(d) \hspace{2cm} r(0)=300}
 % \phantomsubcaption\label{fig:varensvr}
\end{subfigure}
\caption{Plot showing the behaviour of FLI for different Q with different initial conditions r(0) = 12 (top panel) and r(0) = 300 (bottom panel).}\label{fig: FLI_p=6}
\end{figure}

First, we concentrate on the case when the string is initially close to the brane (figure \ref{fig: FLI_p=6}(a),(b)). When $0 < Q < 0.5$, we observe that after some initial transient, the FLI curve becomes vertical indicating the capture of the string. However, unlike p = 5 case, the capture time decreases with charge. When $0.5\le Q \le 1$, the string escapes to infinity. When the string is initially far from the  brane, we 
 find the string escapes to infinity with uniform rate (figure \ref{fig: FLI_p=6}(c),(d)). The approximate linear growth of FLI curve reflects the chaotic motion.
 \newline
 
 Thus, in both charged p=5 and p=6 brane, the chaotic dyanmics do not change when the string starts from a large distance away from the brane!
 \newpage

 Before concluding this section, we comment on the motion of closed string in uncharged p=6 brane. Note that when Q = 0, the corresponding metric becomes a product of 6-dimensional Euclidean space and 4-dimensional Schwarzschild. We obtain the following expression of Hamiltonian:
\begin{equation*}
H = \frac{\pi \alpha^{\prime}}{2} \Big ( \Delta_{+} p_{r}^{2}+\frac{p_{\phi_{1}}^{2}}{r^2} - \frac{p_{t}^{2}}{\Delta_{+}} \Big) + \frac{1}{2 \pi \alpha^{\prime}}n^{2}r^{2}\sin^{2}\phi_{1}    
\end{equation*}
The given system is equivalent to the motion of a string in four dimensional Schwarzschild space-time which has been studied in \cite{frolov_chaotic_1999} where 
 all the three asymptotic modes namely escape to infinity, capture of the string in the event horizon and escape to infinity via back scattering along with an infinite set of unstable periodic orbits  has been reported.  Therein a critical energy threshold has been figured out above which the system turns out to be chaotic.

. %There are three asymptotic modes - escape to infinity, capture of the string in the event horizon and escape to infinity via back scattering along with an infinite set of unstable periodic orbits. We can straightforwardly  demonstrate the chaotic nature by visualizing the string trajectory and the Fast Lyapunov Indicator. First, we set the initial condition condition: r(0) = 10, $p_{r}(0)$ = 0 , $\phi_{1}(0)$ = 0  and
%parameters: n = 1,  $r_{+}$ = 1, E = 6.72
%\newpage
%\begin{figure}[!ht]
% \begin{subfigure}
 %\centering
  %\includegraphics[width=0.51\textwidth]%{radial(p=6_uncharged_r=10).jpg}
%\end{subfigure}
%\hspace{3mm}
%\begin{subfigure}
%\centering
%\includegraphics[width=0.43\textwidth]{FLI(p=6_r=10).jpg}
%\end{subfigure}
%\caption{Plot showing the string orbit and the FLI 
 %in uncharged P=6 black brane}\label{fig: uncharged_p=6 }
%\end{figure}
% From figure \ref{fig: uncharged_p=6 }, note that the string trajectory (left panel) exhibit an oscillatory motion in the beginning, followed by trapping of the string in the event horizon. The corresponding FLI curve (right panel) becomes parallel during the capture which is a manifestation of chaotic motion of the orbit.
\section{Role of winding number}\label{sec: winding}
\begin{figure}[!ht]
 \begin{subfigure}
 \centering
  \includegraphics[width=0.49\textwidth]{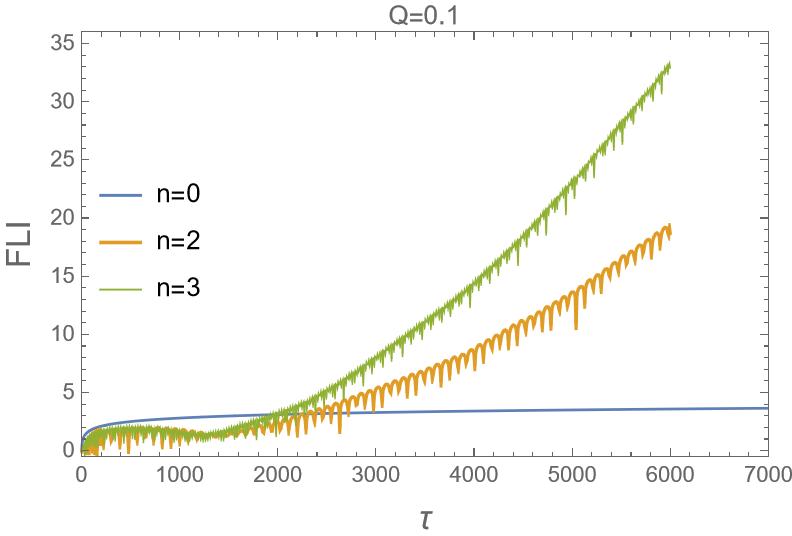}
  \put(-210,136){(a)}
  %\put(-120,106){{\tiny r(0)=11}}
\end{subfigure}
%\hspace{3mm}
\begin{subfigure}
\centering
\includegraphics[width=0.49\textwidth]{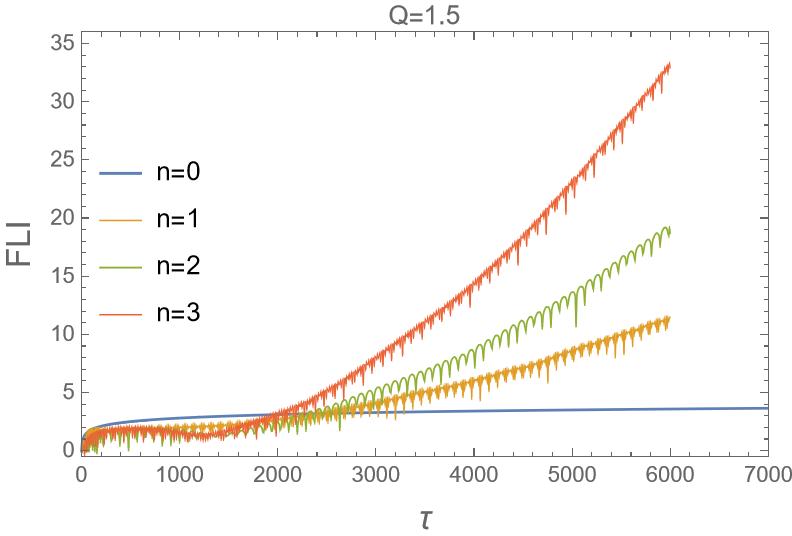}
\put(-210,140){(b)}
%\put(-120,110){{\tiny r(0)=11}}
\end{subfigure}
 \begin{subfigure}
 \centering
  \includegraphics[width=0.49\textwidth]{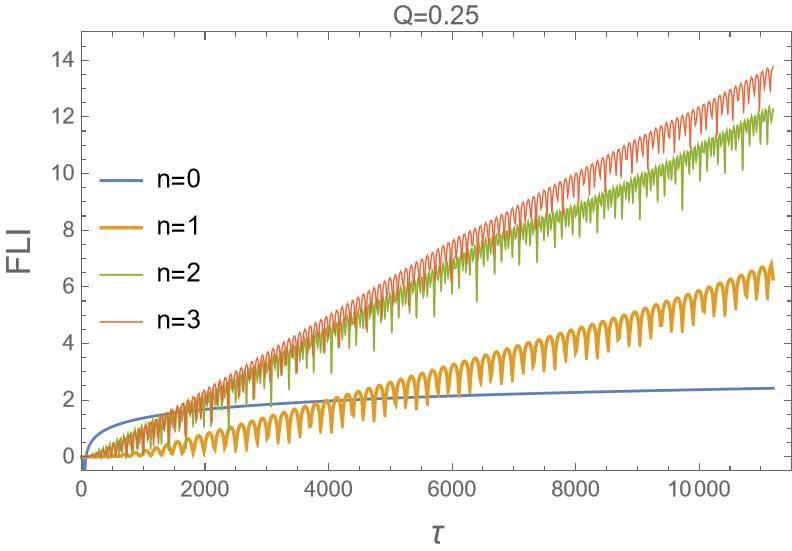}
 \put(-210,140){(c)} 
 %\put(-120,110){{\tiny r(0)=300}}
\end{subfigure}
%\hspace{3mm}
\begin{subfigure}
\centering
\includegraphics[width=0.49\textwidth]{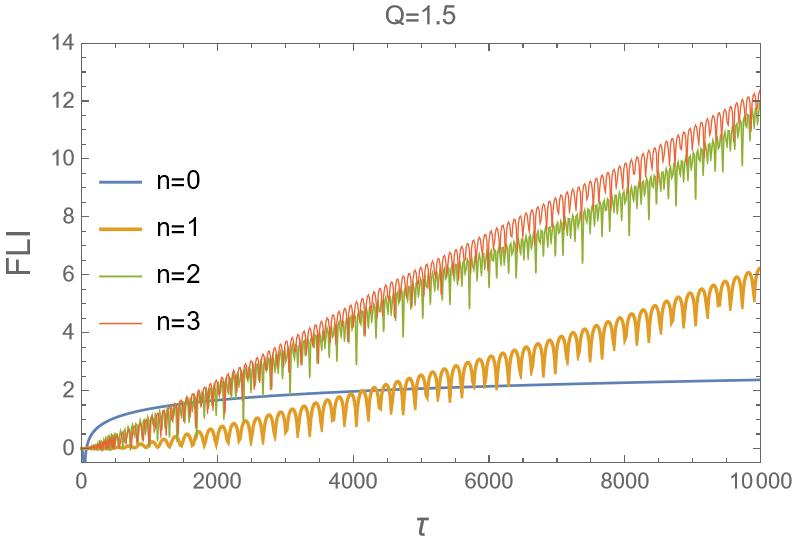}
\put(-210,140){(d)}
%\put(-120,110){{\tiny r(0)=300}}
\end{subfigure}
\caption{Plot showing the FLI for r(0) = 11 (top panel) and r(0) = 300 (bottom panel) for different $n$ and Q in charged p=5 brane. The other parameters are same as in figure \ref{fig: FLI_P=5}.}\label{winding_charged_p=5}
\end{figure}

\begin{figure}[!ht]
 \begin{subfigure}
 \centering
  \includegraphics[width=0.49\textwidth]{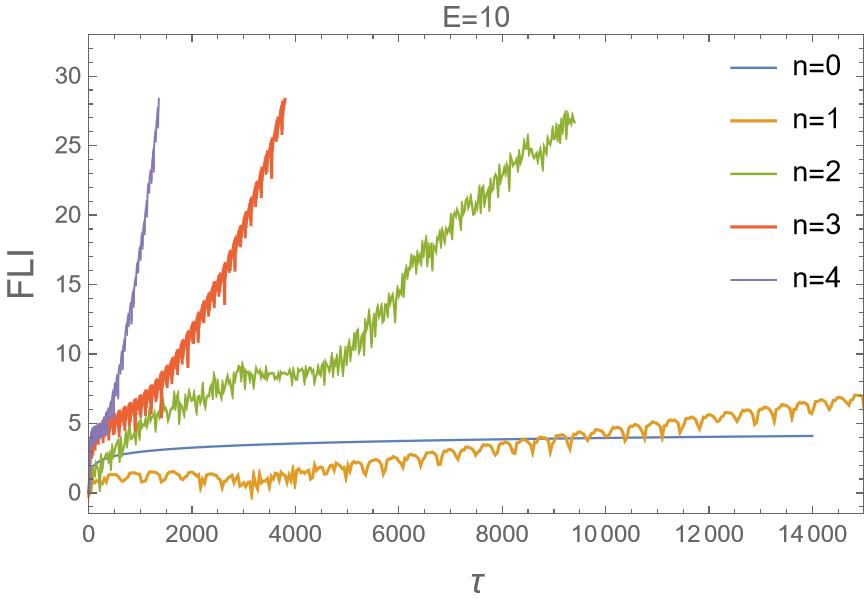}
  \put(-210,143){(a)}
\end{subfigure}
%\hspace{3mm}
\begin{subfigure}
\centering
\includegraphics[width=0.49\textwidth]{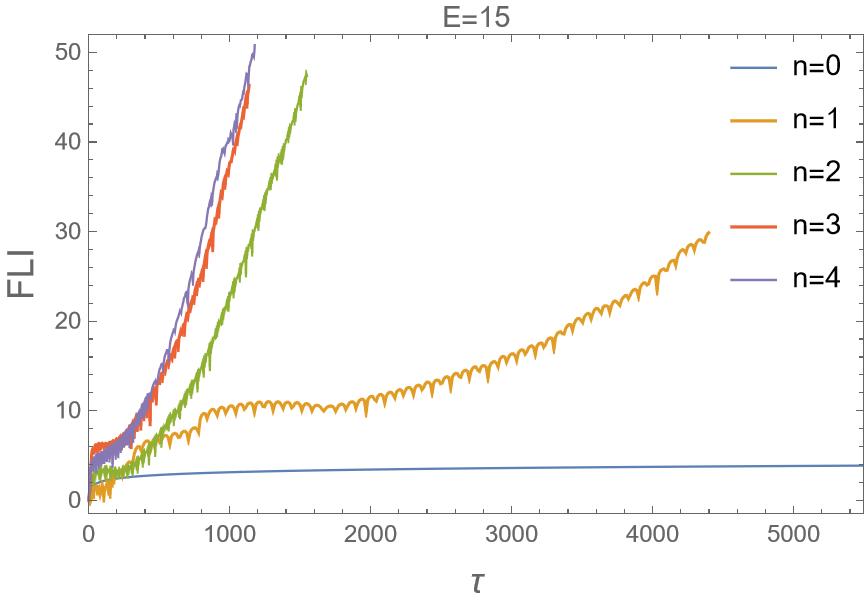}
\put(-210,143){(b)}
\end{subfigure}
 \begin{subfigure}
 \centering
  \includegraphics[width=0.49\textwidth]{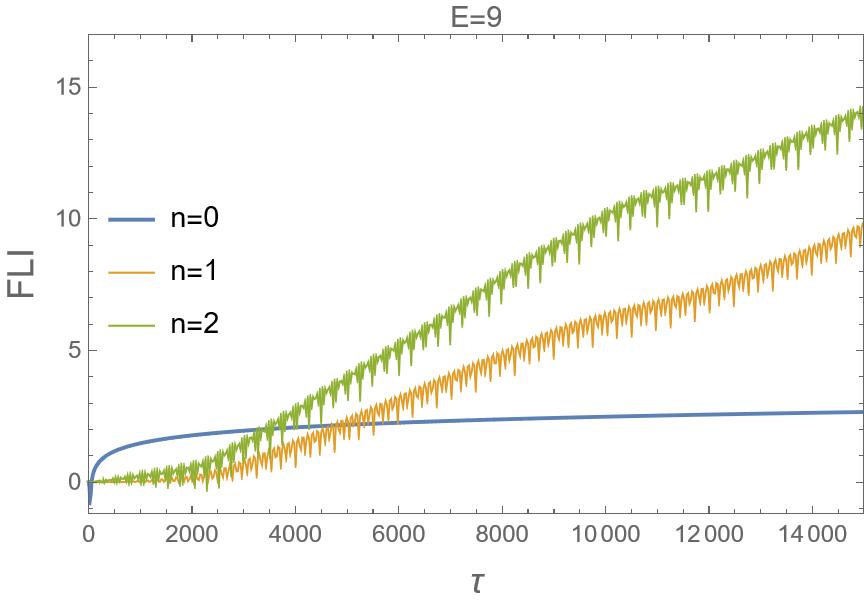}
  \put(-210,135){(c)}
\end{subfigure}
%\hspace{3mm}
\begin{subfigure}
\centering
\includegraphics[width=0.49\textwidth]{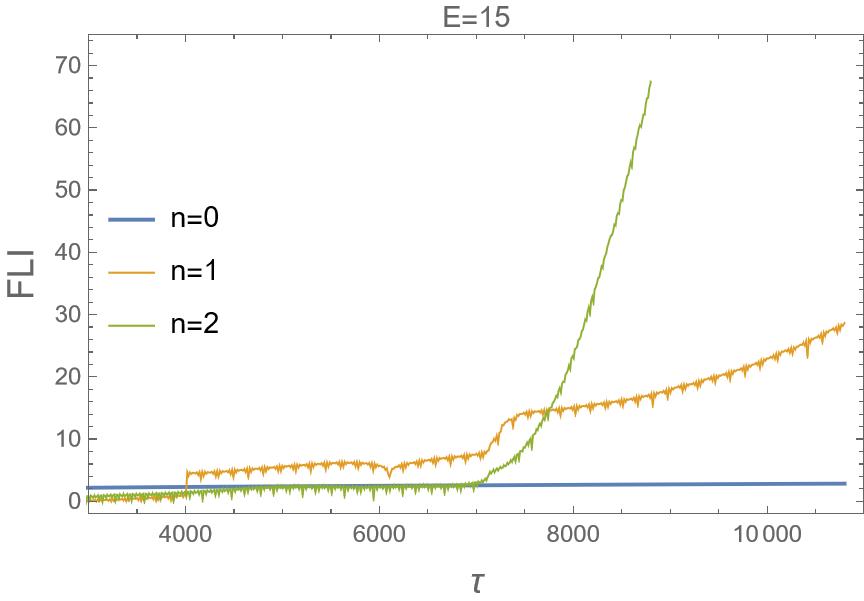}
\put(-210,135){(d)}
\end{subfigure}
\caption{Plot showing the FLI for r(0) = 11 (top panel) and r(0) = 300 (bottom panel) for different $n$ and E in uncharged p=5 brane. The other parameters are same as in figure \ref{fig: FLI(uncharged)} and \ref{fig:FLI(uncharged)2}.}\label{winding_uncharged_p=5}.
\end{figure}

\begin{figure}[!ht]
 \begin{subfigure}
 \centering
  \includegraphics[width=0.49\textwidth]{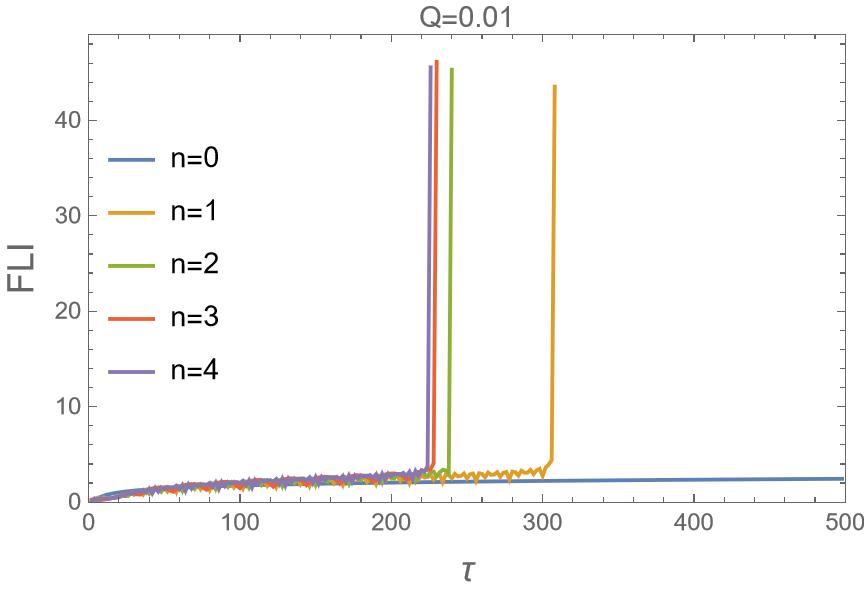}
  \put(-210,140){(a)}
\end{subfigure}
%\hspace{3mm}
\begin{subfigure}
\centering
\includegraphics[width=0.49\textwidth]{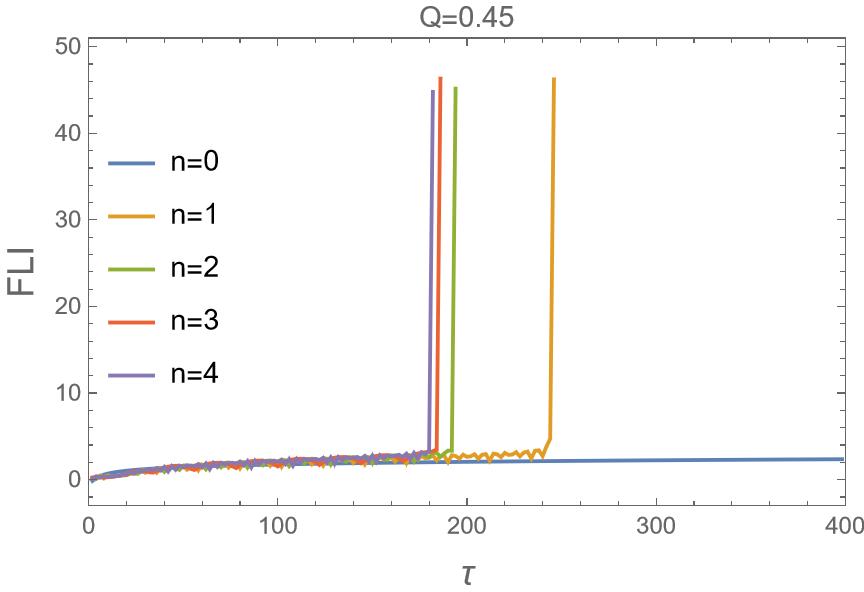}
\put(-210,143){(b)}
\end{subfigure}
 \begin{subfigure}
 \centering
  \includegraphics[width=0.49\textwidth]{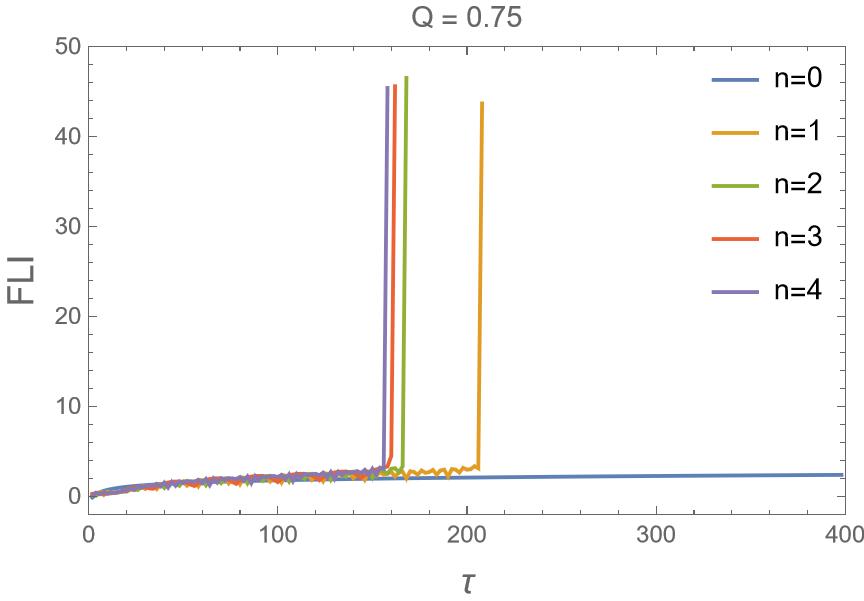}
  \put(-210,143){(c)}
\end{subfigure}
%\hspace{3mm}
\begin{subfigure}
\centering
\includegraphics[width=0.49\textwidth]{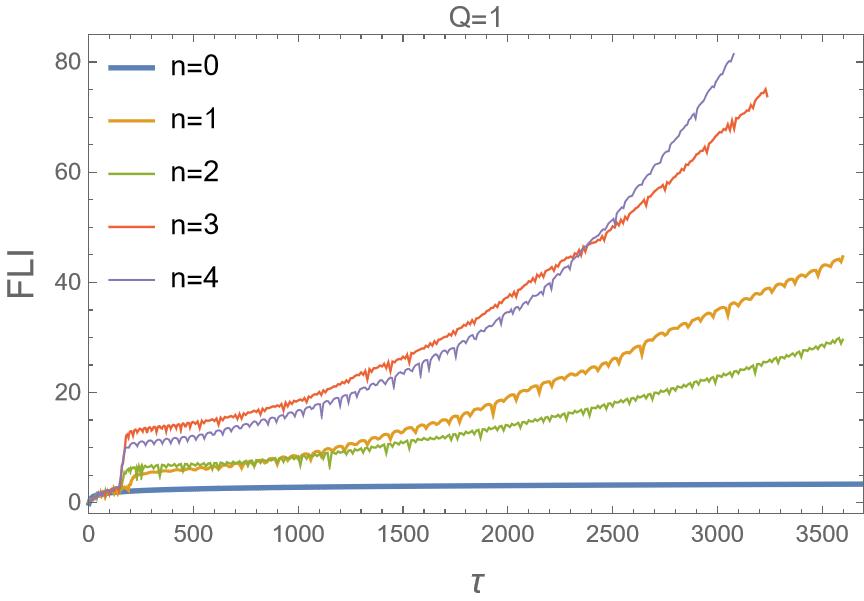}
\put(-210,140){(d)}
\end{subfigure}
\begin{subfigure}
\centering
\includegraphics[width=0.49\textwidth]{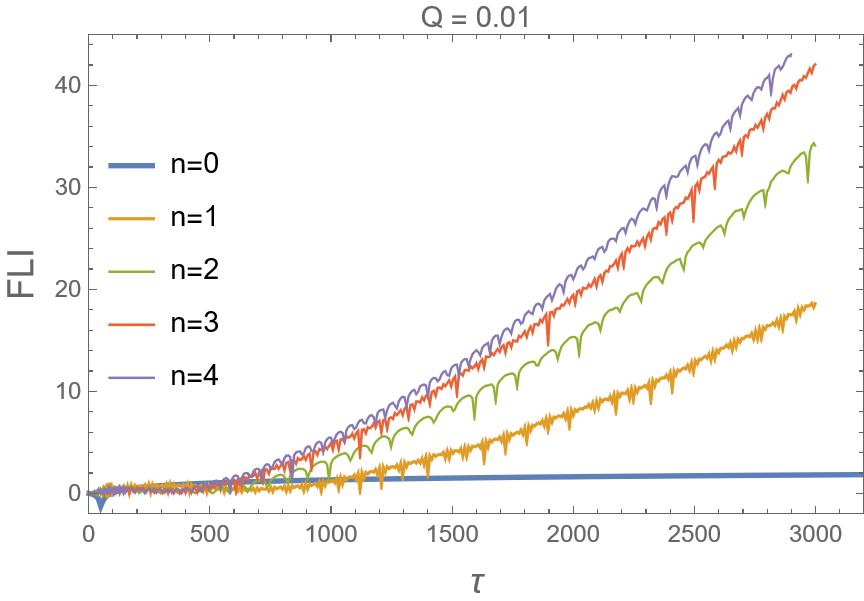}
\put(-210,143){(e)}
\end{subfigure}
 \begin{subfigure}
 \centering
  \includegraphics[width=0.49\textwidth]{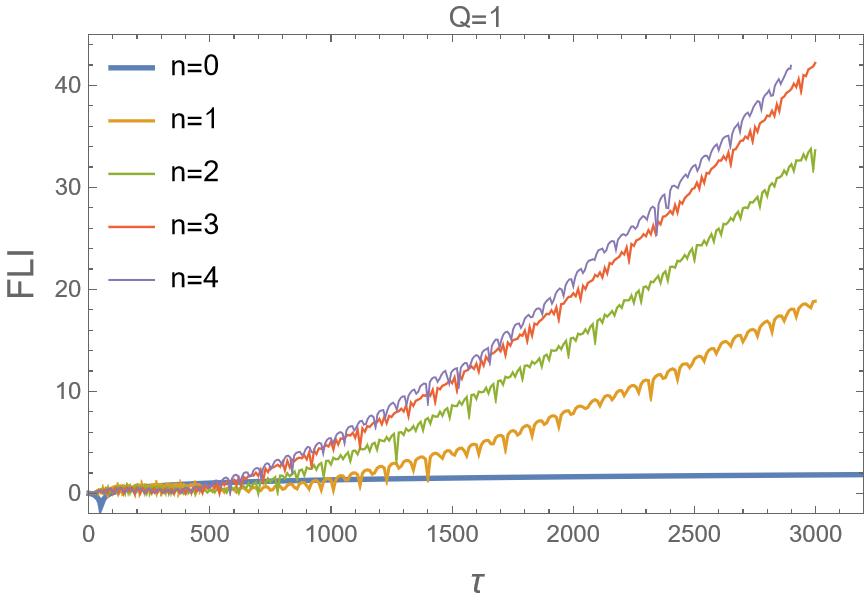}
  \put(-210,143){(f)}
\end{subfigure}
%\hspace{3mm}
\caption{Plot showing the FLI for r(0) = 12 (top and middle panel) and r(0) = 300 (bottom panel) for different $n$ and Q in charged p=6 brane. The other parameters are same as in figure \ref{fig: FLI_p=6}.}\label{winding_charged_p=6}
\end{figure}

We shed some light on the role of winding number in string dynamics. The case $n$=0 corresponds to the point particle. The metric in equation (2.2) has (p+1) translational symmetries due to coordinates $x^\mu (\mu = 0,1,...,p)$ and so has (p+1) constants of motion. By suitable parametrization of the sphere in one higher dimensional sphere, it can be shown that the (D-p-2)-sphere possesses (D-p-2) constants of motion\cite{stepanchuk_nonintegrability_2013}. Therefore, in total, we have $ p + 1 + D - p - 2 + 1 = D$ integrals of motion where the additional integral of motion is coming from the Hamiltonian. Numerically, for $n$=0, the slope of FLI curve is equal to zero (see figure \ref{winding_charged_p=5},\ref{winding_uncharged_p=5},\ref{winding_charged_p=6}) which implies that the corresponding motion is integrable as expected. However, at higher winding numbers, the dynamics is essentially non-integrable which we explain below.\\

First, we consider the charged p=5 brane (figure \ref{winding_charged_p=5}). When $r(0) = 11$ (figure \ref{winding_charged_p=5}(a)) and Q = 0.1, the string escapes to infinity for $n > 1$. However, the string falls very quickly when $n$ = 1 (see figure \ref{fig: FLI_P=5}(a)). With increasing charge (upto Q = 1.5), we 
  see the escape of the string for $n$ = 1(figure \ref{winding_charged_p=5}(b)) also. When  $r(0) = 300$ (figure \ref{winding_charged_p=5}(c),(d)), for both small and large value of Q, the string dynamics at higher $n$ is not very different from what we obtained for $n$ = 1. Note that in all cases, FLI increases linearly and the rate of escape increases with n. %Note that growth of FLI curve are almost same for n = 2,3 and  for n = 4,5.(for both Q=0.1 and Q=1.5).\\
\vspace{3.5mm}
\newpage
Next, we consider the uncharged  p=5 brane (figure \ref{winding_uncharged_p=5}) where we have taken E as a control parameter. When $r(0) = 11$ (figure \ref{winding_uncharged_p=5}(a)) and E = 10, the FLI curve shows quasiperiodic motion for $n$ = 1. The slope (approximately) increases with n reflecting the transition to chaos. For larger energy (E = 15) (figure \ref{winding_uncharged_p=5}(b)), we observe the escape mode for all $n$ $\ge$ 1. When $r(0)$ = 300 (figure \ref{winding_uncharged_p=5}(c),(d)), the string escapes to infinity for both $n$ = 1 and 2.\\
 
Thus, irrespective of where the string is initially located, the distance between two nearby trajectories increases with the winding number.
 \vspace{3.5mm}

Now, we move on to charged p=6 brane (figure \ref{winding_charged_p=6}). First, we consider the case of $r(0) = 12$. For Q = 0.01, 0.45, 0.75 (figure \ref{winding_charged_p=6}(a),(b),(c)), the chaotic dynamics for $n$ = 1 (see also figure \ref{fig: FLI_p=6}(a)) and $n>1$ are not much different except the fact that time of capture of the string decreases with $n$. At the extremal limit (Q = 1), the string escapes to infinity for all $n>1$ (figure \ref{winding_charged_p=6}(d)), however, we do not observe any correlation between the rate of escape and $n$.  When $r(0) = 300$ (figure \ref{winding_charged_p=6}(e),(f)), the string escapes to infinity for all charges and the rate of escape also increases with $n$.
 
\vspace{2.5mm}
\begin{figure}[!ht]
 \begin{subfigure}
 \centering
  \includegraphics[width=0.46\textwidth]{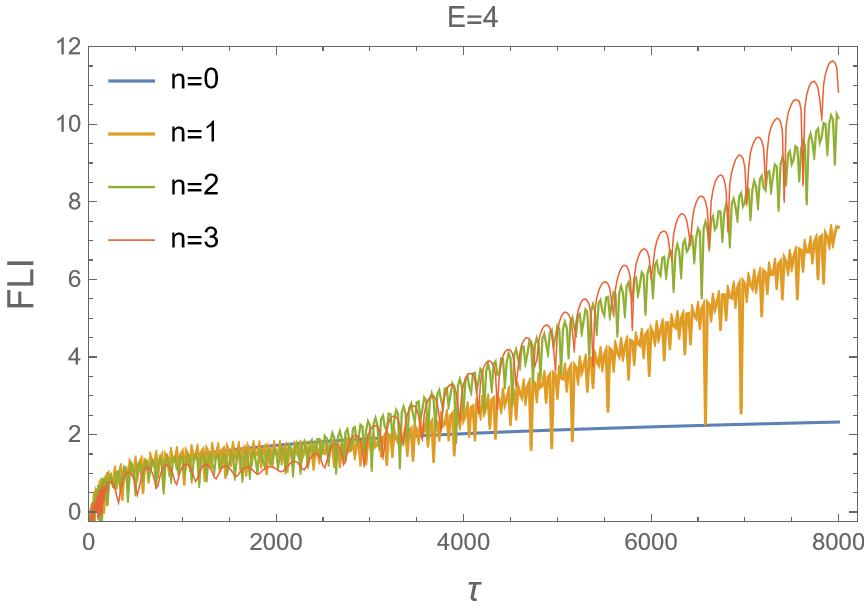}
  \put(-206,136){(a)}
\end{subfigure}
%\hspace{3mm}
\begin{subfigure}
\centering
\includegraphics[width=0.46\textwidth]{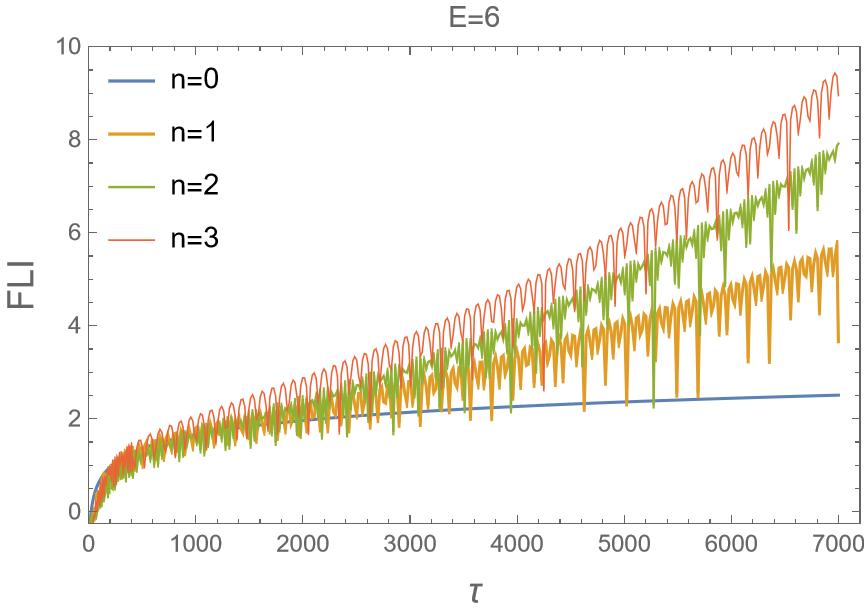}
\put(-206,136){(b)}
\end{subfigure}
 \begin{subfigure}
 \centering
  \includegraphics[width=0.46\textwidth]{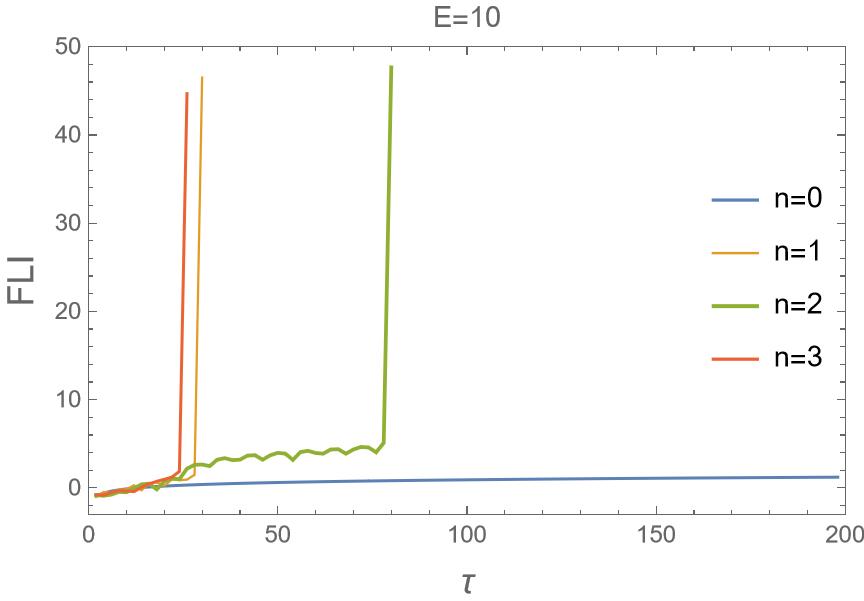}
  \put(-206,136){(c)}
\end{subfigure}
%\hspace{3mm}
\begin{subfigure}
\centering
\includegraphics[width=0.455\textwidth]{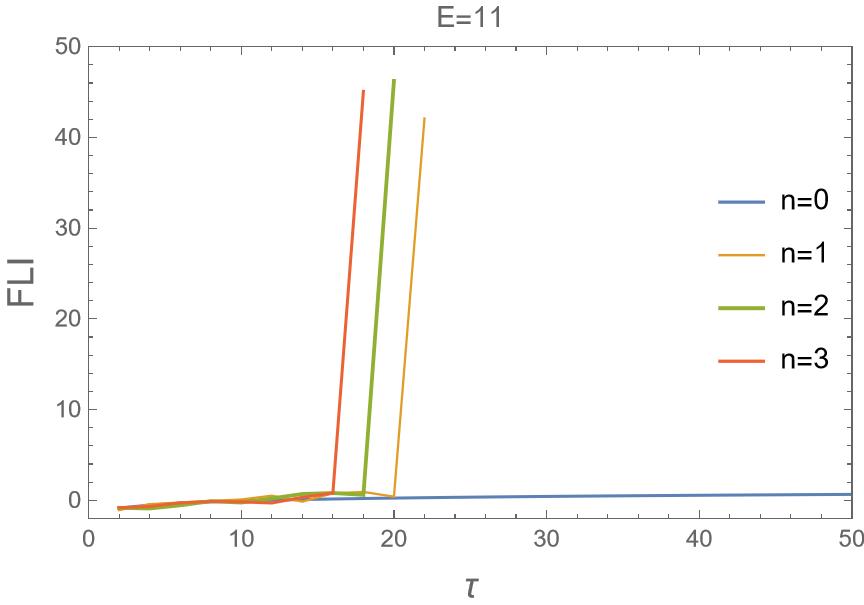}
\put(-206,136){(d)}
\end{subfigure}
\begin{subfigure}
\centering
\includegraphics[width=0.46\textwidth]{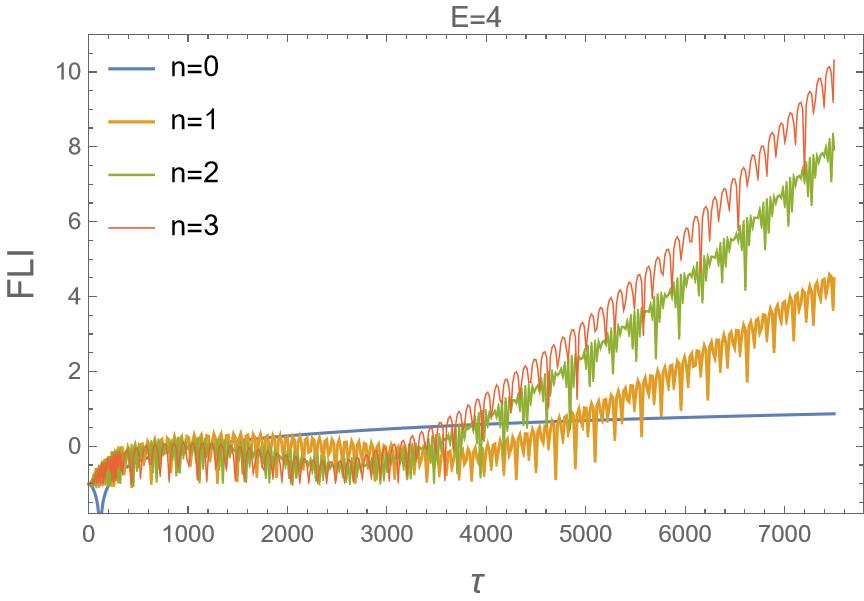}
\put(-206,136){(e)}
\end{subfigure}
\hspace{9mm}
 \begin{subfigure}
 \centering
  \includegraphics[width=0.46\textwidth]{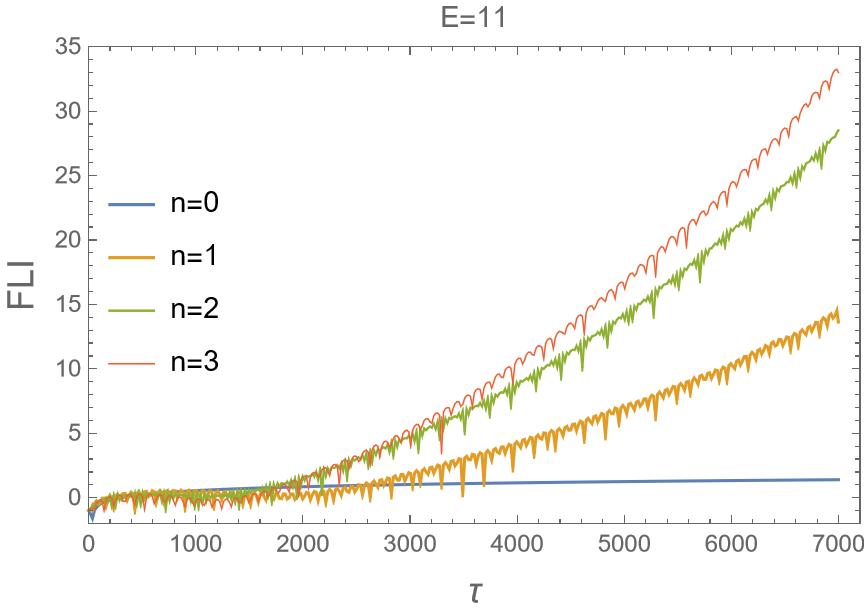}
  \put(-203,136){(f)}
\end{subfigure}
%\hspace{3mm}
\caption{Plot showing the FLI for r(0) = 12 (top and middle panel) and r(0) = 300 (bottom panel) for different n and E in uncharged p=6 brane. The other parameters are same as in figure \ref{winding_charged_p=6}.}\label{winding_uncharged_p=6}
\end{figure}

Finally, we describe the role of $n$ and E in uncharged p=6 brane (figure \ref{winding_uncharged_p=6}) for two different initial locations r(0) = 12 and r(0) = 300. When r(0) = 12 and E = 4,6 we observe linear growth of FLI curve for $n \geq 1$ and the growth increases with n. By increasing energy (E = 10 and E = 11), we observe very quick capture for $n \geq 1$. Note that in the former, the capture time is largest for n = 2 whereas in the latter, the capture time seems to decrease with n. However, for r(0) = 300 (figure \ref{winding_uncharged_p=6}(e),(f)), the string with any non-zero $n$ escapes to infinity for all E, the rate of escape increases with both $n$ and E.

\newpage
 \section{Conclusion and Future directions} \label{Discussion}
 In this work, we have numerically investigated the chaotic behaviour of a circular string in p=5 and p=6 brane and provide sufficient evidence of its chaotic motion. Based on the control parameters of our theory, we summarise the key findings as follows:
 \begin{itemize}
\item Irrespective of the brane-background, the dynamics for $n = 0$ is integrable.
\item  In both charged p=5 and p=6 brane, when the string is initially at a large distance away from the brane, the charge seems to have an insignificant effect on the chaotic dynamics. The effect of winding number is only to increase the rate of escape of the string with $n$  for fixed charge!
\item  For the  charged 5-brane, when the string initially starts near the brane, the capture time of the string increases as a function of charge and eventually at a large charge, the chaotic dynamics change from capture to escape mode. At a higher winding number, the string escapes to infinity, independent of charge and once again, the rate of escape increases with $n$. % although the rate is very small compared to that of the former case.
\item When we disable the charge (Q=0) in 5-brane and study the dynamics by suitably varying energy, independent of where the string is initially located, the dynamics change from quasi periodic to the escape mode  at larger energies. However, the energy at which such a transition occurs that depends on the initial location. With the increase of $n$, the tendency of the string escaping to infinity increases.%which is much more evident at lower energy (E = 10), see figure \ref{fig: uncharged radial p=5}. With the increase of E, the string with any non-zero $n$ escapes to infinity.
\item For both charged and uncharged 6-brane, when the string initially  starts near the brane, our numerics show some non-monotonic behaviour of FLI curve with $n$ for a certain range of parameters.

\item For the charged 6-brane, when $r(0) = 12$, we mostly observe the capture mode of the string, and the capture time decreases with  the increase in  both charge and $n$. %However, when $Q>0.5$, the string escapes to infinity only for $n = 1$. When we visualize the dynamics at higher $n$, again we find the capture time slightly decreases with $n$ for a given charge and is almost the same at Q = 0.7. 
Near the extremal limit, the dynamics change from capture to escape mode for all $n$.%however our numerics show  non-monotonic behaviour of the rate of escape with $n$.
\item For the uncharged 6-brane and when r(0) = 12, we see a transition from escape to capture mode for $n\geq 1$ as we keep increasing control parameter(E). Going far away from the brane (r(0) = 300), we observe only the escape mode of the string. The dynamics at higher winding number follow the same trend, however, the linear growth of FLI seems to increase with $n$.
 \end{itemize}
\vspace{4mm}

In future, we want to explore further the following thought-provoking questions: 
\begin{itemize}
    %\item The comprehensive analytical analysis of the non-integrability of the p-black brane has not yet been conducted. However, we plan to address this gap by employing the Non-Varational Equations(NVE) scheme. 
    %\item %It is a well-known fact that in both classical and quantum mechanical systems, chaos leads to  thermalisation.
    %Note that non-extremal black p-branes are characterized by a finite Hawking temperature. A straightforward computation shows that \cite{ohshima2005comments,duff_black_1996}
%\begin{equation}
       % \beta = 2 \pi \Big[ \frac{2 r_{+} }{7-p} (1- (\frac{r_{-}}{r_{+}})^{7-p})^{\frac{-5+p}{2(7-p)}}     \Big]
   % \end{equation}
\item It was argued in \cite{cubrovic_bound_2019} that for a closed circular string  of winding number $n$ in an AdS black hole, the MSS bound generalizes to 
\begin{equation}\label{eqn:bound2}
\lambda \le 2 \pi k_{B} T n
\end{equation}
 For $n=0$,  the inequality \ref{eqn:bound2} implies that $\lambda = 0$, which is consistent with our findings. We find strong evidence that the dynamics is sensitive to winding number. Also, note that $\beta \sim r_{+}$ for p=5 and $\beta \sim r_{+}\sqrt{1-\frac{r_{-}}{r_{+}}}$ for p=6. This shows that with the increase of charge, Hawking temperature decreases for p=5 and increases for p=6. We do observe that the chaotic nature depends on charge/temperature at least when the string is initially close to the brane. This motivates us to do a comprehensive  study of the near-horizon dynamics together with the relation \ref{eqn:bound2} in the context of the p-black brane. We plan to address this gap by employing the NVE scheme \cite{ruiz1999differential,morales-ruizIntegrabilityHamiltonianSystems2007} in future work similar to \cite{cubrovic_bound_2019}.
\item Expanding the directions of chaos, it would be interesting to study the scrambling properties of our system. \cite{Susskind:2018tei,prihadi_chaos_2023}

\item In the future for lower and higher dimensions of branes, we would like to do a detailed study of the generic p-branes involving a method of consistent truncation and dimensional reduction.
\item It would be interesting to explore the chaotic behaviour of string and point particle in intersecting non-extremal p-branes of \cite{miaoCompleteIntersectingNonExtreme2004}.
%\item \EDIT{Using a similar approach, it would be interesting to explore and  check the chaotic indicators in the generic intersecting non-extremal p-branes of }\cite{miaoCompleteIntersectingNonExtreme2004}
\end{itemize}
\section*{Numerical accuracy and error}
\begin{figure}[!ht]
    \centering
    \includegraphics[width=0.5\textwidth]{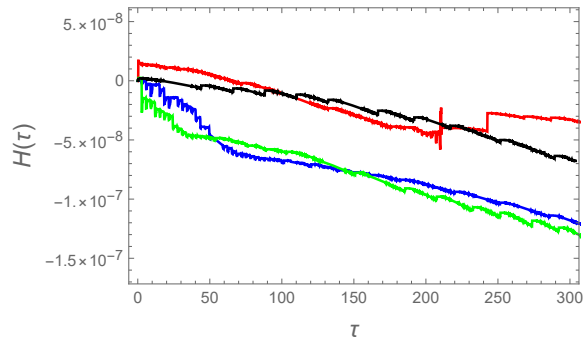}
    \caption{The evolution of H in charged p=5(blue), uncharged p=5(green), charged p=6(red) and uncharged p=6(black) brane.}
    \label{fig: error}
\end{figure}

In this paper, we use the \texttt{Projection} method of \texttt{NDSolve} routine of \texttt{Mathematica} to solve the equations of motion. Note that we are dealing with nonlinear differential equations and  one has to keep track of the constraint H = 0 at every integration step. We have checked this numerical accuracy and  found that the maximum possible error (with in a reasonable computation time) is of the order of $10^{-6}$ (figure \ref{fig: error}).
\newpage
\section*{Acknowledgments}
The authors would also like to thank Manoranjan Samal for valuable comments on the manuscript.

%\section{Some examples}
%\label{sec:intro}

%For internal references use label-refs: see section~\ref{sec:intro}.
%Bibliographic citations can be done with "cite": refs.~\cite{a,b,c}.
%When possible, align equations on the equal sign. The package
%\texttt{amsmath} is already loaded. See \eqref{eq:x}.
%\begin{equation}
%\label{eq:x}
%\begin{aligned}
%x &= 1 \,,
%\qquad
%y = 2 \,,
%\\
%z &= 3 \,.
%\end{aligned}
%\end{equation}

\bibliographystyle{JHEP}
%\bibliography{biblio.bib}
%\bibliography{chaoticstring.bib}
\bibliography{chaoticstring}
% Bibliography

%% [A] Recommended: using JHEP.bst file
%% \bibliographystyle{JHEP}
%% \bibliography{biblio.bib}

%% or
%% [B] Manual formatting (see below)
%% (i) We suggest to always provide author, title and journal data or doi:
%% in short all the informations that clearly identify a document.
%% (ii) please avoid comments such as "For a review'', "For some examples",
%% "and references therein" or move them in the text. In general, please leave only references in the bibliography and move all
%% accessory text in footnotes.
%% (iii) Also, please have only one work for each \bibitem.

%\begin{thebibliography}{99}

%\bibitem{a}
%Author,
%\emph{Title},
%\emph{J. Abbrev.} {\bf vol} (year) pg.

%\bibitem{b}
%Author,
%\emph{Title},
%arxiv:1234.5678.

%\bibitem{c}
%Author,
%\emph{Title},
%Publisher (year).

%\end{thebibliography}
\end{document}